\documentclass[12pt, reqno, oneside,amsart]{article}
\usepackage[
vmarginratio=1:1, hmarginratio=1:1,
left=1.2in,right=1.2in,
top=1in,bottom=1.2in,includeheadfoot
						]{geometry} 
\usepackage{etoolbox}
\usepackage{titlesec}
\usepackage{titletoc}

\usepackage{amsmath}

\titleformat{\section} 
	{
	\filcenter
	\normalsize
	\scshape
	} 
{\thesection.}       
{0.3em}                  
{}

\titlespacing{\section}
{0pt}                  
{20pt}                  
{13pt}                  

\titleformat{\subsection} 
[runin]                 
{
\bfseries
} 
{\thesubsection.}       
{0.3em}                  
{}                     

\titlespacing{\subsection}
{0pt}                  
{20pt}                  
{5pt}                  

\titleformat{\subsubsection} 
[hang]
{
\normalsize
\bfseries
} 
{\thesubsubsection.}    
{0.3em}                
{}                     

\titlespacing{\subsubsection}
{0pt}                  
{25pt}		        
{15pt}                  

\usepackage{amsthm}	

\usepackage{mathrsfs}  
\def\scr{\mathscr}

\usepackage{cite}
\usepackage{graphicx}
\usepackage{amsfonts}
\usepackage{amssymb}

\usepackage[dvipsnames]{xcolor}	

\usepackage{stmaryrd}	

\input epsf

\usepackage[shortlabels]{enumitem}

\usepackage{hyperref}
\hypersetup{
    colorlinks=true,
    anchorcolor=red,
    urlcolor=magenta,
    citecolor=red,
    linkcolor=cyan}

\def\a{\alpha}

\def\s{\sigma}

\def\la{\lambda}

\def\rm{\mathrm}
\def\cal{\mathcal}

\def\pa{\partial}

\def\be{\begin{equation}}
\def\ee{\end{equation}}
\def\br{\begin{eqnarray}}

\def\er{\end{eqnarray}}
\def\bsub{\begin{subequations}}
\def\esub{\end{subequations}}

\def\Oint{O^{(\rm{int})}_{2}}
\def\Oi{O^{(\rm{int})}}

\def\lor{\colon\!\!}
\def\ror{\! \colon\!}

\usepackage{dsfont}		

\usepackage{simplewick}    

\usepackage{braket}

\def\no{\noindent}
\def\p{\partial}

\title{Correlation functions of composite Ramond fields  in deformed  D1-D5 orbifold SCFT$_2$ }

\usepackage{authblk}

\author[1]{\normalsize A.A. Lima\thanks{andrealves.fis@gmail.com}}
\author[1]{\normalsize G.M. Sotkov\thanks{gsotkov@gmail.com}}
\author[2]{\normalsize M. Stanishkov\thanks{marian@inrne.bas.bg}}

\affil[1]{\textit{\footnotesize Department of Physics, Federal University of Esp\'irito Santo, 29075-900, Vit\'oria, Brazil}}
\affil[2]{\textit{\footnotesize Institute for Nuclear Research and Nuclear Energy, Bulgarian Academy of Sciences, 1784 Sofia, Bulgaria}}

\begin{document}

\date{}

\begin{titlepage}

\maketitle

\begin{abstract}

We study  two  families of  composite twisted Ramond fields (made by products of two operators) in the $\cal {N}=(4,4)$ supersymmetric  D1-D5 SCFT$_2$  deformed by a marginal modulus operator away from its $(T^4)^N/ S_N$ free orbifold  point.  We construct  the large-$N$ contributions to the  four-point functions with two composite  operators and two deformation fields. These functions allow us to derive short-distance OPE limits and  to calculate  the anomalous dimensions of the composite operators. 
We  demonstrate that
 one can distinguish  two sets of composite Ramond states with twists $m_1$ and  $m_2$:  protected states,  for which $m_1+m_2=N$, and  ``lifted''  states for which $m_1+m_2<N$. The latter require an appropriate renormalisation. We also derive the leading order   corrections to their two-point functions, and  to their three-point functions with the deformation operator.

{\footnotesize 
\bigskip
\noindent
\textbf{Keywords:}

\noindent
Microstate black hole geometries; Symmetric $\cal {N}=4$ SUSY orbifold CFTs; Correlation functions.
}

\end{abstract}

\pagenumbering{gobble}

\end{titlepage}

\pagenumbering{arabic}

\tableofcontents

\section{Introduction} \label{secIntroduction}

The scalar moduli deformation of the symmetric orbifold  $(T^4)^N/S_N$ gives rise to  a particular  two-dimensional  $\mathcal {N}=(4,4)$  superconformal theory with  central charge $c=6N$, which for large values of $N$  provides a  fuzzball \cite{Mathur:2005zp} description of  certain five-dimensional  extremal supersymmetric black holes.
Their type IIB superstring counterparts are bound states of the D1-D5 brane system
(see e.g.~\cite{David:2002wn}, and \cite{Warner:2019jll} for a more recent review), which gave the first microscopical account of the Bekenstein-Hawking entropy \cite{Strominger:1996sh}. 
There is strong evidence 
\cite{Alday:2006nd,Kanitscheider:2006zf,Kanitscheider:2007wq,Skenderis:2006ah,Skenderis:2008qn,Taylor:2007hs}
that  appropriate coherent superpositions of twisted Ramond states (and certain products of them)
reproduce the  ``microstate geometries'' holographically dual to the  semiclassical IIB supergravity 2-charge horizonless nonsingular solutions of AdS$_3 \times S^3 \times T^4$ type. Similar statements hold for the microstates of the more realistic near-extremal 3-charge $1/8$-BPS black holes, the so-called D1-D5-P system, which can be realized as  appropriate tensor products of the (left-right non-symmetric) descendants of twisted Ramond ground states of the  same D1-D5 orbifold SCFT$_2$ \cite{Giusto:2012yz,Lunin:2012gp,Bena:2015bea}. A more complete description of the quantum properties of such SUSY black holes requires further investigation of the spectra  of conformal dimensions of composite fields, the construction of their correlation functions, and analysis of their renormalization as an effect of the interaction introduced by  the marginal perturbation away from the free orbifold point.

Despite numerous  results and achievements  
\cite{Burrington:2012yq,Galliani:2017jlg,Burrington:2017jhh,Tormo:2018fnt,Bena:2018mpb,Bena:2019azk,Bena:2020yii,Bombini:2017sge,Guo:2019pzk,Guo:2019ady,Belin:2019rba,Guo:2020gxm, Giusto:2020mup},
 the super-conformal data concerning the effects of the interaction in the deformed D1-D5 SCFT$_2$  remains incomplete. As we have demonstrated in a recent paper \cite{Lima:2020boh}, the simplest  R-charged  twisted Ramond fields  $R^{\pm}_n(z,\bar z)$ get renormalized, i.e.~their conformal dimensions and  certain structure constants  acquire corrections  in the perturbed theory.
 It is then natural to address the question of whether  the simplest composite states  $R^{\pm}_{m_1}R^{\pm}_{m_2}(0)$, made by a product of two Ramond fields with  twists $m_1$ and $m_2$,  are BPS-protected or should be renormalized. If renormalization occurs to some fields, what are, then, the conditions  defining classes of  ``protected'' and ``lifted'' Ramond states in the deformed theory?

The answer to the above questions  requires  the explicit construction  of the large-$N$  contributions to the four-point correlation functions  involving two composite Ramond fields and two deformation operators.
This is what we compute in the present paper, using the `covering surface technique' \cite{Lunin:2000yv} together with the `stress-tensor method' \cite{Dixon:1986qv,Arutyunov:1997gt,Pakman:2009ab,Pakman:2009mi}.
We compute the four-point function at the 
genus-zero order of the genus expansion \cite{Lunin:2000yv} for large $N$, and we find that the four-point function decomposes into a sum of ``connected'' and ``partially-disconnected'' parts.
Our result allows us to examine certain  short-distance limits, and to compute  the structure constants  as well as  the conformal dimensions of  the  specific non-BPS descendants of  twisted fields present in these OPEs.

Once we have the explicit form of the four-point functions, integrating over the positions of the interaction operators yields the correction to the conformal dimensions 
of $R^{\pm}_{m_1}R^{\pm}_{m_2}$, to second order in perturbation theory.
The nature of the composite operators crucially depends on the properties of  the twists $m_1$ and $m_2$ of their components. We demonstrate that operators with $m_1+m_2=N$ form a family of \emph{protected states}, whose   conformal dimensions  remain the same as in the free orbifold point because the correction vanishes.  The remaining composite fields, with $m_1+m_2<N$, suffer from  certain UV divergences and do require an appropriate renormalisation; as a result, their  conformal dimensions get corrected.

\section{Symmetric orbifold D1-D5 SCFT$_2$} \label{orbi}

In this paper we are concerned with a symmetric orbifold model $(T^4)^N/S_N$ where $T^4$ is a four dimensional torus
and $S_N$ is the corresponding symmetric group. This SCFT$_2$ orbifold model is considered as a ``free orbifold  point'' of the D1-D5 system (see for example \cite{David:2002wn,Seiberg:1999xz}).

The theory contains $4N$ free scalar fields $X^i_I$, with $i=1, \cdots ,4$ and $I=1, \cdots, N$, and $4N$ free fermions $\psi^{i}_I$, with total central charge $c_{\rm{orb}}=6N$. The $N$ copies of the fields are identified by the action of the symmetric group: $X^i_I(e^{2\pi\i}z,e^{-2\pi\i}\bar z)=X^i_{g(I)}(z,\bar z)$, where $g \in S_N$. These boundary conditions are realized by twist fields $\sigma_g(z)$, which give a representation of $S_N$. 
For example $\sigma_{(1\cdots n)}$ imposes the cyclic permutations of the fields corresponding to the cycle $(1\cdots n)$,
\be
X^i_1\rightarrow X^i_2 \rightarrow \cdots \rightarrow X^i_n \rightarrow X_1,
\ee
and similarly for the fermions. 
We denote by $\s_n$ the twist field corresponding to the conjugacy class obtained by summing over the orbits of $(1\cdots n)$,
 \be
 \s_n = {1\over \scr S_n}\sum_{h\in S_N} \sigma_{h^{-1}(1\cdots n)h} ,
 	\label{DefsSninv}
 \ee
 with $\scr S_n(N)$ a combinatorial factor ensuring the normalization of the two-point function of the $S_N$-invariant operators,
 \be
 \langle \s_n (z,\bar z) \s_m(0) \rangle = \frac{\delta_{mn}}{|z |^{4\Delta^\s_n}} .	\label{2ptfunsingtw}
 \ee
 We call attention for a notational convention that we use throughout the paper:
 a twist index without brackets, like  in $\s_n$, indicates a \emph{sum over conjugacy classes} of cycles of length $n$, as in the r.h.s.~of Eq.(\ref{DefsSninv}). A twist index with brackets, like in $\s_{(n)}$, indicates one single twist corresponding to a specific permutation cycle $(n)$, of length $n$; e.g.~$\s_{(2)}$ is a short notation for $\s_{(12)}$ or $\s_{(37)}$ or $\s_{(15)}$, etc. 
The holomorphic and anti-holomorphic dimensions of $\s_n(z,\bar z)$, $\Delta^\sigma_n$ and $\tilde \Delta^\s_n$ respectively, and of any non-$S_N$-invariant twist field $\s_{(n)}(z,\bar z)$,  are 
\be\label{twistdim}
\Delta^{\sigma}_n={1\over 4}\left( n-{1\over n} \right) =  \tilde \Delta^\s_n .
\ee

We further pair the $4N$ real scalar fields into complex bosons $X^a_I$ and $X^{a\dagger}_I$, $a=1,2$. The Majorana fermions can also be combined into complex fermions and then bosonized by the use of $2N$ new free scalars: $\psi_I^a=e^{i\phi_I^a}$, $\psi_I^{a\dagger}=e^{-i\phi_I^a}$.
The holomorphic sector possesses $\cal N=4$ superconformal symmetry, generated by the stress-energy tensor $T(z)$, the SU(2) currents $J^i(z)$, ($i=1,2,3$) and the supercurrents $G^a(z)$, $\hat G^a(z)$ ($a=1,2$). These currents are expressed in terms of the free fields. For example, the stress tensor is given by
\be\label{stres}
\begin{split}
T(z) = - \frac{1}{2}\lim_{w \to z} \sum_{a =1}^2 &\sum_{I=1}^N \Bigg( \pa X^a_I(z) \pa X^{a\dagger}_I(w)
+\pa\phi^a_I(z)\pa\phi^a_I(w)+{6\over (z-w)^2} \Bigg) ,
\end{split}
\ee
for the $J^3$ current of the SU(2) algebra defining the conserved R-charge we have
\be\label{j3}
J^3(z)={i\over 2}\sum_{I=1}^N (\p\phi^1_I+\p\phi^2_I) (z) ,
\ee
while anti-holomorphic  currents are built from $\bar \pa X^a_I(\bar z)$, $\bar \pa X^{a\dagger}_I(\bar z)$ and $\tilde \phi^a_I(\bar z)$.

In the orbifold model, one has to consider distinct sectors: Ramond, Neveu-Schwarz (NS)  and twisted, representing different boundary conditions for the constituent free fermions and bosons.
Ground state twisted Ramond fields (those of dimension ${c/ 24}$) have a simple realization in terms of the free fields,
\be\label{micro1}
\begin{split}
 R_n^{\pm}(z) = \frac{1}{\scr S_n} 
 	\sum_{h\in S_N} 
	\exp \Big( \pm {i\over 2n} \sum_{I=1}^n [\phi_{h(I)}^1(z) +\phi_{h(I)}^2 (z) ]
	\Big)
	\sigma_{h^{-1}(1\cdots n)h}(z) .
\end{split}\ee
From this holomorphic field, we can define $R^\pm_n(z,\bar z) \equiv R^\pm_n(z) \tilde R^\pm_n(\bar z)$. Eq.(\ref{micro1}) is an explicitly $S_N$-invariant construction, normalized by the combinatorial overall factor. The holomorphic dimension and R-charge are
\be\label{microdim}
\Delta^R_{n}=\tfrac{1}{4} n , \qquad j^3=\pm \tfrac{1}{2}.
\ee
By construction, the $R^\pm_n$ are doublets of the SU(2) R-symmetry algebra and singlets of the global  SU(2)$_1$ algebra.%
	\footnote{%
	There are other neutral composite Ramond fields made by the intrinsically R-neutral single-cycle fields which form a doublet under the ``internal'' SU(2)$_1$ inherited from the target-space $T^4$ symmetry group $\rm{SO}(4) = \rm{SU}(2)_1 \times \rm{SU}(2)_2$. We will not consider these fields here.
	}
In this paper we will be  actually interested in composite fields made of products of two of the twisted  Ramond fields (\ref{micro1}). 
 More precisely, we will consider two types of composite Ramond fields, 
 \be
 R_{m_1}^{\pm}R_{m_2}^{\pm}(z, \bar z) , \qquad R_{m_1}^{\mp}R_{m_2}^{\pm}(z, \bar z) ,	\label{ComRampres}
 \ee
 which are, respectively, charged and neutral under R-symmetry --- under the action of the ``isospin'' SU(2) algebra, these products of  $j={1\over2}$ representations form a triplet with $j^3=  -1, 0, +1$, respectively given by 
\be
R_{m_1}^{-}R_{m_2}^{-} , 
\quad 
\tfrac{1}{\sqrt{2}}  \big( R_{m_1}^{+}R_{m_2}^{-}  + R_{m_1}^{-}R_{m_2}^{+} \big), 
\quad 
R_{m_1}^{+}R_{m_2}^{+}
\ee
and  a singlet  $\frac{1}{\sqrt{2}} \left( R_{m_1}^{+}R_{m_2}^{-}  -  R_{m_1}^{-}R_{m_2}^{+} \right)$ as well.
Composite Ramond fields play a role in the microstate description of  the near-horizon and the interior of certain five-dimensional extremal supersymmetric black holes (or black rings) which can be realized semi-classically as AdS$_3\times S^3\times T^4$ solutions of  type IIB supergravity. Within the AdS/CFT correspondence, they permit  a particular dual holographic description in terms of a definite SCFT$_2$ model realized as a marginal deformation of  the symmetric orbifold $(T^4)^N/ S_N$ SCFT with (large) central charge $c=6N$ \cite{Maldacena:1998bw,Seiberg:1999xz}, cf. also \cite{David:2002wn}.

The Hilbert space of the orbifold theory can be organized as the direct sum
$\cal H_{\rm{orb}} =  \oplus_{[g]} \cal H_{[g]}$ of Hilbert spaces $\cal H_{[g]}$ containing the states invariant under elements in the conjugacy class $[g]$ of a $g \in S_N$.
Conjugacy classes of $S_N$ are equivalent to partitions of $N$, i.e.~to sets $\{k_j\}$ of $N$ integers such that $\sum_{j = 1}^N j k_j = N$, which define the cycle structure of the elements in $[g]$, 
\be
g = (1)^{k_1} (2)^{k_2} \cdots (N)^{k_N} , \qquad \sum_{k = 1}^N j k_j = N ;	\label{ConjClassPar}
\ee
here $(n_j)^{k_j}$ is a composition of $k_j$ disjoint cycles of length $n_j$. The untwisted sector corresponds to $[ 1]$, the conjugacy class of the unity, for which $k_1 = N$ and $k_{j\neq 1} = 0$, while the Hilbert space where the composite field $R^\pm_{m_1}R^\pm_{m_2}$ lives corresponds to the equivalence class of $(1)^{N-m_1-m_2} (m_1)(m_2)$.
In the operator language, we construct double-cycle operators such as $R^\pm_{m_1}R^\pm_{m_2}(z,\bar z)$ with double-cycle twist operators defined by a `normal-ordered' product of two single-cycle twists \cite{Roumpedakis:2018tdb}:
\be	\label{NormComtwsdef}
\begin{split}
&\lor \s_{m_1}\s_{m_2} \ror  \equiv
 \frac{1}{\scr C_{m_1m_2}}
 \sum_{h\in S_N} \sigma_{h^{-1}(1,\cdots ,m_1)h} \sigma_{h^{-1}(m_1 +1 , \cdots,  m_1 + m_2)h} ,
\end{split}
\ee
This normal-ordering amounts to eliminating from the r.h.s.~products of cycles with overlapping elements (and then summing over the orbits). For example, 
$$
\lor \s_{2} \s_{3} \ror \supset \s_{(12)} \s_{(345)} + \s_{(13)} \s_{(542)} + \s_{(64)} \s_{(312)} + \cdots
$$
while terms like $\s_{(12)} \s_{(234)} = \s_{(1342)}$ are absent from the r.h.s.
The normalization factor appearing in the definition (\ref{NormComtwsdef}) is
\be
\scr C_{m_1m_2} = \scr S_{m_1} \scr S_{m_2} .	\label{nomrCm1m2}
\ee
This indeed ensures normalization because the composite two-point function factorizes into a product of single-cycle two-point functions:
\begin{align*}
&	 \frac{1}{\scr C_{n_1n_2}} \frac{1}{\scr C_{m_1m_2}}
		\sum_{g , h \in S_N}
	\Big\langle
		\sigma_{g^{-1}(1,\cdots ,n_1)g} \sigma_{g^{-1}(n_1 +1 , \cdots,  n_1 + n_2)g}
		\sigma_{h^{-1}(1,\cdots ,m_1)h} \sigma_{h^{-1}(m_1 +1 , \cdots,  m_1 + m_2)h}
	\Big\rangle	
\\
	&= \frac{\scr S_{m_1}^2 \scr S_{m_2}^2}{\scr C_{m_1m_2}^2}
		\langle
		\sigma_{m_1} \sigma_{m_1} 
		\rangle
		\langle
		\sigma_{m_2} \sigma_{m_2} 
		\rangle
\end{align*}
where we have been rather schematic;  using (\ref{2ptfunsingtw}) and (\ref{nomrCm1m2}),
\begin{align}
\big\langle \lor \s_{n_1}\s_{n_2} \ror (z,\bar z) \lor \s_{m_1}\s_{m_2} \ror (0) \big\rangle
	= 
		\frac{1}{|z|^{2(\Delta^\s_{m_1} + \Delta^\s_{m_2})}}	
\end{align}	
from which see that the dimension of $\lor \s_{m_1} \s_{m_2}\ror$ is $\Delta^\s_{m_1,m_2} = \tilde \Delta^\s_{m_1,m_2} = \Delta^\s_{m_1} + \Delta^\s_{m_2}$.

Double-cycle twisted fields are built by dressing the double-cycle twist operators; in particular,
\be\label{micro1comp}
\begin{split}
&R_{m_1}^{\pm} R_{m_2}^{\pm}(z) 
	= 
\\
&	\frac{1}{\scr C_{m_1m_2}}
	\sum_{h\in S_N} 
	\exp \Bigg[ 
	\pm{i\over 2m_1}\sum_{I=1}^{m_1} \big( \phi_{h(I)}^1 +\phi_{h(I)}^2 \big) 
	\pm{i\over 2m_2}\sum_{I=m_1+1}^{m_1+m_2} \big( \phi_{h(I)}^1 +\phi_{h(I)}^2 \big)
	\Bigg]
\\	&\qquad\qquad\qquad
	\times
	\sigma_{h^{-1}(1\cdots m_1)h} 
	\sigma_{h^{-1}(m_1+1\cdots m_1+m_2)h} ,
\end{split}
\ee
to be compared with (\ref{micro1}). These operators have R-charge $j^3 = \pm \frac{1}{2}$, and holomorphic dimension $\Delta^R_{m_1+m_2}$.
Let us make a remark that (\ref{micro1comp}) always involves two Ramond operators, corresponding to the two first cycles in the equivalence class $(m_1)(m_2)(1)^{N-m_1- m_2}$. 
The other $N-m_1-m_2$ trivial cycles correspond to (untwisted) NS vacua.
When one of the Ramond cycles becomes trivial, say $m_2 = 1$, the corresponding field $R^\pm_{(1)}$ becomes the spin field appearing in the untwisted Ramond sector.
We use the notation $R^\pm_{m_1} R^\pm_{m_2}$ instead of, say, $R^\pm_{m_1,m_2}$, precisely to emphasize this point.

\section{Correlation functions of composite Ramond fields}	\label{SectTwoPointFuncts}

We are interested in 
the two- and three-point functions of composite Ramond fields in the marginally perturbed theory,
\be
S_{\rm{def}}(\lambda)=S_{\rm orb}+ \lambda \int d^2u \, \Oint (u,\bar {u}) \label{def-cft}
\ee
where $\la$ is a dimensionless coupling constant, and the deformation operator $\Oint$ is an $S_N$-invariant SU(2) scalar, preserving $\cal {N}=(4,4)$ supersymmetry. Its explicit form
\be\label{interaction}
\Oint (u,\bar {u})=  \left( \hat G^1_{-1/2}\bar G^2_{-1/2}- G^2_{-1/2}\bar{\hat G}^1_{-1/2}\right) O_{2} (u,\bar{u})+ c.c. 
\ee
is a sum of descendants of the twist-two NS chiral field $O_{2}$ with conformal dimensions $\Delta_2+ \tilde \Delta_2 = 1$ and SU(2) charges $j_3=\frac12=\tilde j_3$. 
See e.g.~\cite{Burrington:2012yq}.

The conformal dimension of the composite operator $R^\pm_{m_1} R^\pm_{m_2}(z, \bar z)$, 
at the free orbifold point, is given by the sum of the dimensions of its constituents, i.e.
\[
(\Delta^R_{m_1,m_2} , \tilde \Delta^R_{m_1,m_2}) =   \Big( \frac{m_1+m_2}{4} , \ \frac{m_1+m_2}{4} \Big).
\]
The first nontrivial correction to the two-point function 
\be
\big\langle R_{m_1}^- R_{m_2}^- (\infty) \ R_{m_2}^+ R_{m_1}^+(0) \big\rangle_{\lambda}	\label{2ptRpp}
\ee
 appears at second order in perturbation theory,
\be
\begin{split}
\frac{\lambda^2}{2}  \int \! d^2z_2 \int & \! d^2z_3 \, 
	\Big\langle R_{m_1}^-R_{m_2}^-(z_1, \bar z_1) \, \Oint (z_2, \bar z_2)  \,
	 \Oint(z_3, \bar z_3) \, R_{m_2}^+ R_{m_1}^+ (z_4, \bar z_4) \Big\rangle .
\end{split}
\label{second-cor}
\ee
Conformal invariance  fixes the form of the  four-point functions up to an arbitrary function $G(u,\bar{u})=G(u)\bar{G}(\bar u)$ of the anharmonic ratio $u = z_{12}z_{34} /  z_{13}z_{24}$ and its complex conjugate $\bar u$,
\be
\begin{split}
&\big\langle R_{m_1}^- R_{m_2}^- (z_1,\bar z_1) \Oint(z_2, \bar z_2) \Oint (z_3, \bar z_3) R_{m_2}^+ R_{m_1}^+ (z_4, \bar z_4) \big\rangle
		=\frac{|z_{14}|^{4-m_1-m_2}}{|z_{13}z_{24}|^{4}}G(u,\bar u). 
\end{split}
\label{4-point-R-1}
\ee
One can further make a suitable change of variables and factorize the integral (\ref{second-cor}). As a result we get for the first nontrivial correction to the two-point function,
\be\label{correct}
  \frac{\lambda^2\pi}{|z_{14}|^{m_1+m_2}} \log \frac{\Lambda}{ |z_{14}|}  \int \! d^2u \, G (u,\bar u),
\ee
where $\Lambda$ is an ultraviolet cutoff, and we have used SL(2,$\mathbb C$) invariance to fix three points in the correlation function, so that
\be
 G(u,\bar u) = \big \langle R_{m_1}^-R_{m_2}^-(\infty) \Oint(1) \Oint(u, \bar u) R_{m_2}^+R_{m_1}^+(0) \big \rangle.
	\label{gu}
\ee


\subsection{Connected and disconnected functions.}	\label{SectConnDisconnFunc}

The $S_N$-invariant function (\ref{gu}) is a sum over the group orbits,
\be \label{micro1G}
\begin{split}
G(u,\bar u) =  & \sum_{S_N} 
		\Big \langle 
		R^-_{h^{-1}_\infty (m_1) h_\infty}
		R^-_{h^{-1}_\infty (m_2) h_\infty} (\infty) 
		\Oi_{h_1^{-1} (2) h_1}(1) 
	\\
	&\qquad\qquad
	\times
		\Oi_{h_u^{-1} (2) h_u}(u, \bar u) 
		R^+_{h^{-1}_0(m_2)h_0}
		R^+_{h^{-1}_0 (m_1) h_0}(0) \Big \rangle
\end{split}\ee
summation being over every $h_\infty, h_1, h_u, h_0 \in S_N$.
Each individual term in this sum corresponds to one of the possible individual permutations resulting from the composition of the six permutation cycles $(n_i)$,  ordered by (the radial order of) the points $z_i$ where the twists $\s_{(n_i)}(z_i)$ are located. Following \cite{Pakman:2009zz}, we will denote the permutation of the twist field $\s_{(n_i)}(z_i)$ by the cycle $(n_i)_{z_i}$,  labeled by a position index.
The cycles in Eq.(\ref{micro1G}) are accordingly denoted as  $(m_1)_\infty (m_2)_\infty (2)_1 (2)_u (m_2)_0 (m_1)_0$.  
(Imposing an ordering is crucial, since $S_N$ is non-abelian.)
Every permutation contributing to the sum (\ref{micro1G}) must satisfy the condition
\be	\label{mm2mmcond}
(m_1)_\infty (m_2)_\infty (2)_1 (2)_u (m_2)_0 (m_1)_0  = 1 ,
\ee
otherwise the correlation function vanishes.
Some of the correlators in the r.h.s.~of Eq.(\ref{micro1G}) factorize in different ways, and some will be completely connected.

A term in the sum (\ref{micro1G}) will be (fully) connected when
one of the elements of $(2)_1 = (k, \ell)$, say $k$, overlaps with $(m_1)_\infty$, and the other element, $\ell$, overlaps with $(m_2)_\infty$. Because of (\ref{mm2mmcond}), a similar overlap will happen for $(2)_u$, $(m_1)_0$ and $(m_2)_0$.
In this case, there is always a number
\be
{\bf s}_c = m_1 + m_2 \label{scoloc}
\ee
of different elements entering the permutation $(m_1)_\infty (m_2)_\infty (2)_1 (2)_u (m_2)_0 (m_1)_0$.

A four-point function in the sum (\ref{micro1G}) can factorize in three qualitatively different ways which do not vanish. 
Factorization depends on the existence of cycles commuting with all the others, which is regulated by the different possibilities of overlapping  the elements of the cycles $(2)_1$ and $(2)_u$ with the other cycles, since $(m_1)$ and $(m_2)$ are always disconnected. 
The first possibility is that  $(2)_1$ and $(2)_u$ commute with every Ramond-operator cycles.
Then the four-point function splits into 
\be	\label{vacuumbubble}
 \big\langle \Oi_{(2)}(1) \Oi_{(2)^{-1}} (u, \bar u) \big\rangle 
 \;
 \big \langle R_{(m_1)}^- R_{(m_2)}^-(\infty) R_{(m_1)}^+ R_{(m_2)}^+(0) \big \rangle 
\ee
with $(m_1)_\infty (m_2)_\infty (m_1)_0 (m_2)_0 =1$.
In this case, the integral (\ref{second-cor}) is over the ``vacuum bubbles'' $\big\langle \Oi_{(2)}(1) \Oi_{(2)^{-1}} (u, \bar u) \big\rangle$, which diverge. These divergences are natural in perturbation theory, and can be eliminated by proper normalization of the correlation functions,
\be
 \frac{\big \langle R_{m_1}^-R_{m_2}^-(\infty) \Oint(1) \Oint(u, \bar u) R_{m_1}^+R_{m_2}^+(0) \big \rangle_\la}{\langle \mathds 1 \rangle_\la} . \label{DiscoBubbles}
 \ee
 We will assume this normalization from now on but omit the $\langle \mathds 1 \rangle_\la$, so terms like (\ref{DiscoBubbles}) are henceforth excluded from (\ref{gu}).

The other two possibilities are of a very different nature.
If the pairs of cycles with lengths $m_1$ or $m_2$ commute with the other cycles, than we have the factorizations
\bsub	\label{factzdab}	\begin{align}
\begin{split}
&\big\langle R_{(m_2)}^- (\infty) R_{(m_2)^{-1}}^+(0) \big\rangle  
\times
	 \big\langle  R_{(m_1)}^-(\infty)   \Oi_{(2)} (1) \Oi_{(2)} (u,\bar u)R_{(m_1)}^+(0) \big\rangle , \label{factzd1}
\end{split}\end{align}
and
\begin{align}
\begin{split}	 
&\big\langle R_{(m_1)}^- (\infty) R_{(m_1)^{-1}}^+(0) \big\rangle   \;
\times
 \big\langle  R_{(m_2)}^-(\infty)   \Oi_{(2)} (1) \Oi_{(2)} (u,\bar u)R_{(m_2)}^+(0) \big\rangle , \label{factzd2}
\end{split}	 
\end{align}\esub
where  $(m_1)_\infty (2)_1 (2)_u (m_1)_0 = 1$ in (\ref{factzd1}), and $(m_2)_\infty(2)_1 (2)_u (m_2)_\infty =1$ in (\ref{factzd2}), so as to satisfy (\ref{mm2mmcond}).
Note that, if a term in (\ref{factzdab}) factorizes further, it has the form (\ref{vacuumbubble}) and is canceled by (\ref{DiscoBubbles}).
We are going to call functions like (\ref{factzdab}) \emph{`partially disconnected'} (and, hereafter, when we say just `disconnected function' we implicitly mean `partially disconnected').
Denote by $k,\ell$ the elements of  $(2)_1 = (k, \ell)_1$, then look at the permutation $(m_1)_\infty (m_2)_\infty (2)_1$. 
There are two qualitatively different ways in which the factorizations (\ref{factzdab}) happen, as follows.
\begin{enumerate}[label={\itshape \arabic*)}]
\item \label{Factr1}
Only \emph{one} of the elements of $(2)_1$, say $k$, overlaps with $(m_2)_\infty$, while the other element, $\ell$, does \emph{not} overlap with \emph{any} of the $(m_1)$ nor the $(m_2)$ cycles. 
This gives a factorization (\ref{factzd2}).

\noindent
A factorization (\ref{factzd1}) happens when one of the elements of $(2)_1$, say $k$, overlaps with $(m_1)_\infty$, and the other element, $\ell$, does not overlap with any of the $(m_1)$ nor the $(m_2)$ cycles.
In any case, there is always a number
\be
{\bf s} = m_1 + m_2 + 1	\label{scolo1}
\ee
of \emph{distinct} elements entering the permutation (\ref{mm2mmcond}).

\item \label{Factr2}
\emph{Both} $k$ and $\ell$  overlap with $(m_2)_\infty$ or, instead, both overlap with $(m_1)_\infty$. These possibilities are mutually exclusive, since $(m_1)_\infty$ and $(m_2)_\infty$ do not share elements.


Concerning the number of different elements appearing in the permutation, in Case \ref{Factr2} there are two different situations. For simplicity, let us drop indices and  call the ``non-factorized'' permutation simply $(m)_\infty(2)_1 (2)_u(m)_0$. We can use $S_N$ symmetry to fix $(m)_\infty = (1,2,3, \cdots, m)$ and $(2)_1 = (1 , \ell)$. 

\bigskip

	\begin{enumerate}[label={\itshape 2\alph*)}]
	
	\item \label{Factr2a}
	In the generic case, we have $\ell \neq 2$ and $\ell \neq m$. Then the permutation splits into
	$
	(1,2, \cdots, \ell , \cdots,  m)_\infty (1, \ell)_1 = (1, \cdots, \ell-1)(\ell, \cdots, m) .
	$
	Hence there is a number $m$ of distinct elements which should also appear in $(2)_u(m)_0$ so that $(m)_\infty(2)_1 (2)_u (m)_0 = 1$. Counting these elements together with the other ``factorized'' ones, we find
	\be
	{\bf s} = m_1 + m_2  \label{scolo2}
	\ee
	distinct elements entering the r.h.s.~of (\ref{mm2mmcond}).
	
	\item \label{Factr2b}
	However, if $\ell = 2$ or $\ell = m$, then the permutation
	$(1,2,3, \cdots, m)_\infty(1,\ell)_1$ collapses to a cycle with length $m-1$.
	For example, if $\ell = m$, then
	$$
	(1,2,3, \cdots, m)_\infty (1,m)_1 = (1,2,\cdots,m-1).
	$$
	Now the permutation $(2)_u (m)_0$, which must equal the inverse cycle, can accommodate \emph{one more distinct element}, which is \emph{not} in $\{1,2,\cdots,m\}$, because 
	$$
	(r,1)_u (r,m-1, \cdots, 2,1)_0 = (m-1, \cdots, 2, 1)
	$$
	for \emph{any} $r \in [1,N]$, not only for $r = m$.
	There are, therefore, $m+1$ elements entering the ``non-factorized'' permutation,
	hence ${\bf s} = m_1 + m_2 + 1$ distinct elements entering the permutation (\ref{mm2mmcond}), the same number (\ref{scolo1}).

	\end{enumerate}

\end{enumerate}

The sum over orbits preserves the cycle structure of factorized functions, hence the function (\ref{micro1G}), normalized as (\ref{DiscoBubbles}), splits into three terms:
\be
G(u,\bar u) = G_c (u, \bar u) + G_{m_1}  (u, \bar u) + G_{m_2}  (u, \bar u),	\label{Guu}
\ee
where
\begin{align}
\begin{split}
&G_c(u, \bar u) 
= \big \langle  R_{m_1}^- R_{m_2}^- (\infty) \, \Oint(1) \, \Oint(u, \bar u) \,  R_{m_1}^+ R_{m_1}^+  (0) \big \rangle_{\rm{conn}} \label{Gcconnec}
\end{split}
\end{align}
and
\begin{align}
G_{m_1}(u,\bar u) &=   \big\langle  R_{m_1}^-(\infty)  \, \Oint (1) \, \Oint (u, \bar u) \, R_{m_1}^+(0) \big\rangle
\label{G1factorzd}
\\
G_{m_2} (u,\bar u) &=  \big\langle  R_{m_2}^-(\infty)   \Oint (1) \Oint (u, \bar u)R_{m_2}^+(0) \big\rangle 
\label{G2factorzd}
\end{align}
(Note the twist indices without parenthesis,  indicating that each of the correlators are (multiple) sums over orbits.) 
We emphasize that all correlators are normalized as (\ref{DiscoBubbles}), and the `conn' in (\ref{Gcconnec}) indicates that there is no factorization of the composite operators. 
The Ramond two-point functions in (\ref{factzdab}) have disappeared because of the normalization (\ref{micro1}) --- after summing  over orbits, the factored two-point functions are $\big\langle (R_{m_p}^\pm)^\dagger R_{m_p}^\pm \big\rangle = 1$. 
The functions $G_{m_1}$ and $G_{m_2}$ are four-point functions of \emph{non-composite} operators, and have been considered in \cite{Lima:2020boh}.
The integral (\ref{correct}) over these terms \emph{does not vanish}, hence renormalization of the Ramond fields is required to cancel the logarithmic divergence in Eq.(\ref{correct}).
We will return to this point later.
For most of the remaining of this section, we focus on function $G_c$.

\subsection{Large-$N$ limit.}

We are interested in theories with $N \gg 1$. 
To find the $N$-dependence of the correlation functions, 
following \cite{Pakman:2009zz}, we can first organize the sum (\ref{Gcconnec}) according to the conjugacy classes $\a$ of the symmetric group.
This  is very convenient because $S_N$-invariance implies that every term belonging to the same conjugacy class $\a$ must give the same result. 
For large $N$ it is further convenient to separate permutations inside a class according to the number $\bf s$ of \emph{distinct} `active' elements, i.e. elements which undergo non-trivial permutations.%
	\footnote{%
	For example, the permutation $(259)(3)(14)(7)$ has five `active' elements: 1, 2, 4, 5 and 9.}
Then (we omit the `conn' hereafter)
\be \label{Gsumoverclassandcolor}
\begin{split}
G_c(u,\bar u) = \sum_{{\bf s}} \sum_{\a_{\bf s}}  C_{{\bf s}, \a_{\bf s}} (N) 
&	\Big\langle 
		R_{g^{\a_{\bf s}}_\infty }^-R_{g'^{\a_{\bf s}}_\infty}^-(\infty) 
		\Oi_{g^{\a_{\bf s}}_1} (1) 
		\Oi_{g^{\a_{\bf s}}_u}(u, \bar u) 
		R_{g'^{\a_{\bf s}}_0}^+R_{g^{\a_{\bf s}}_0}^+(0) \Big\rangle,
\end{split}		
\ee
where $\a_{\bf s}$ is the set of permutations belonging to class $\a = \cup_{\bf s} \a_{\bf s}$ and involving $\bf s$ distinct active elements. The individual permutation appearing in the twists in (\ref{Gsumoverclassandcolor}),
$g^{\a_{\bf s}}_\infty g'^{\a_{\bf s}}_\infty g^{\a_{\bf s}}_1 g^{\a_{\bf s}}_u g'^{\a_{\bf s}}_0 g^{\a_{\bf s}}_0 \in {\a_{\bf s}}$,
is one arbitrary representative of ${\a_{\bf s}}$.

The numerical symmetry factor $C_{{\bf s}, \a_{\bf s}} (N) $ counts the number of elements in $\a_{\bf s}$ times normalization factors $\scr S_r(N)$, $r = 1,\cdots,6$, present in the definition of $S_N$-invariant fields. Note there is one factor of $1/\scr S_r(N)$ for every cycle $(n_r)$ entering the permutation, including the two cycles in each composite operator, since $\scr C_{m_1 m_2} = \scr S_{m_1} \scr S_{m_2}$.
The symmetry factor $C_{{\bf s}, \a_{\bf s}} (N) $ can be computed exactly with some combinatorics similar to what is done in \cite{Lunin:2000yv,Pakman:2009zz} and, when $N \gg n_r$, as we can ignore overlappings, its large-$N$ dependence can be found with a very simple argument due to \cite{Lunin:2000yv} --- there are $\bf s$ different elements entering the permutation, which can be chosen in $N^{\bf s}$ ways; meanwhile, 
$1/\scr S_{n_r} \sim N^{-\frac{1}{2} n_r}$, the $\frac12$ in the exponent due to $\scr S_{n_r}$ being a normalization factor for the two-point function $\langle \s_{n_r} \s_{n_r} \rangle$ (where there are $n_r$ distinct active elements). Hence
\be \label{CoefflarN}
C_{{\a_{\bf s}, \bf s}} (N) = N^{{\bf s} - \frac{1}{2} \sum_{r=1}^q n_r } \left[ \varpi(n_r)  +  \rm{O}(1/N) \right] ,
\ee
where the function $\varpi(n_r)$ does not depend on $N$.
It turns out that (even the exact) result does not depend on the class $\a$, only on $\bf s$.
Here, for the function (\ref{Gsumoverclassandcolor}), $q = 6$ and $n_1 = m_1 = n_6$, $n_2 = m_2 = n_5$, $n_3 = 2 = n_4$.
But we have used a notation such that the result holds for a $q$-point function involving $q$ single twists of length $n_r$, $r = 1,\cdots,q$. In particular, it holds for the partially disconnected four-point functions (which have $q = 4$).

The exponent of $N$ in (\ref{CoefflarN}) can be recast into an interesting form using the Riemann-Hurwitz formula
\be
 {\bf g} = \frac{1}{2} \sum_{r=1}^q (n_r - 1) - {\bf s} + 1
  \label{RiemHurwForm}
 \ee
 for the genus $\bf g$ of a surface $\Sigma$ which is a ramified covering of the sphere possessing $\bf s$ sheets and $q$ ramification points with ramification orders%
	\footnote{%
	In the standard definition \cite{lando2002ramified,lando2013graphs,cavalieri2015riemann}, the `order' of the ramification points is $n_r - 1$, not $n_r$, but we make this abuse of language for convenience.
	Recall that the ramification points $\{t_1, \cdots , t_q\} \in \Sigma$ are points on the covering surface $\Sigma$, whose image under the covering map $z(t)$ are the branching points $\{z_1 , \cdots, z_s\} \in S^2_{\rm{cover}}$ of the base sphere. In general,  $q \geq s$.
	}
$n_r$ \cite{Lunin:2000yv,Pakman:2009zz}.
 In terms of $\bf g$, 
\be
C_{{\bf s}} (N) 
\equiv  C_{{\bf g}} (N)
	\sim N^{- {\bf g} - \frac{1}{2} q + 1} \Big( \varpi  +  \rm{O}(1/N) \Big)  .	\label{largNCg}
\ee
Using $\Sigma$ as a `covering surface' of the base sphere is the standard way of calculating correlation functions in the orbifold theory \cite{Lunin:2001pw}, as we will do later.
Eq.(\ref{largNCg}) shows that the leading contribution at large $N$ comes from surfaces of genus zero. The specific power of $N^{- \frac{1}{2} q + 1}$ for ${\bf g} = 0$ then depends on the number $q$ of ramification points of the covering surface.

We can thus replace the sum over $\bf s$ in (\ref{Gsumoverclassandcolor}) by a sum over genera,
\begin{align}
& G_c(u,\bar u) 
	= \sum_{{\bf g} = 0}^{{\bf g}_{\max}}  C_{{\bf g} } (N) 
	\sum_{\a_{\bf g}}
	\Big\langle 
		R_{g^{\a_{\bf g}}_\infty }^-R_{g'^{\a_{\bf g}}_\infty}^-(\infty) 
		\Oi_{g^{\a_{\bf g}}_1} (1) 
		\Oi_{g^{\a_{\bf g}}_u}(u, \bar u) 
		R_{g'^{\a_{\bf g}}_0}^+R_{g^{\a_{\bf g}}_0}^+(0) \Big\rangle
\nonumber
\\
	&\quad
	= \frac{\varpi + \rm{O}(1/N)}{ N^{ \frac{1}{2} q - 1} }
	\sum_{\frak a = \frak1}^{{\bf H}_c}
	\Big\langle 
		R_{(n_1)_\infty}^- R_{ (n_2)_\infty }^-(\infty) 
		\Oi_{(n_3)_1} (1) 
		\Oi_{(n_4)_u}(u, \bar u) 
		R_{(n_5)_0}^+R_{(n_6)_0}^+(0) \Big\rangle_{\frak a}
\nonumber\\
	&\qquad 
	+ \text{higher-genera}	
 \label{GsumGen} 
\end{align}
In (\ref{GsumGen}) we have kept only terms at leading-order in $1/N$, and corresponding to ${\bf g} = 0$. This is still a sum over ${\bf H}_c$ conjugacy classes satisfying ${\bf g} = 0$, and we label the representative functions for each of these classes by an index $\frak a = \frak1, \frak2, \frak3, \cdots$. The number ${\bf H}_c$ is a `Hurwitz number' \cite{Pakman:2009ab,Pakman:2009zz}, which will be determined in two different ways in \S\ref{SectCovMaps} and in Appendix \ref{AppHurNum}. 

It is well-known that there is a fundamental interplay between the permutation cycles dictating the monodromy of the correlation functions and the properties of the corresponding covering surfaces, addressed by Hurwitz Theory \cite{Lunin:2000yv,Pakman:2009ab,Pakman:2009zz}.%
	\footnote{See \cite{lando2002ramified,lando2013graphs}, and \cite{cavalieri2015riemann} for a friendly introduction.}
The number $\bf s$ of sheets of $\Sigma$ is equal to the number of distinct elements entering non-trivially in the permutations  of the twisted correlation function. 
We have seen in \S\ref{SectConnDisconnFunc} that the different types of disconnected functions have different $\bf s$. For the factorized \emph{two}-point functions in (\ref{factzdab}), we have $q = 2$ ramification points of order $m = (m_1$ or $m_2)$, and ${\bf s} = m$; hence Eq.(\ref{RiemHurwForm}) gives ${\bf g}_2 = 0$ and Eq.(\ref{largNCg}) shows that these functions go to a constant $\sim N^{0}$ at large $N$.

Meanwhile, the corresponding factorized non-composite four-point functions in (\ref{factzdab}) have $q = 4$ ramification points of orders $m, 2,2,m$, with $m = (m_2$ or $m_1$); their ${\bf s}$ depends on the types of factorization: for Types \ref{Factr1} and \ref{Factr2b}, ${\bf s} = m + 1$ hence ${\bf g} = 0$, giving a dependence $\sim N^{-1}$. For Type \ref{Factr2a}, however, ${\bf s} = m$, hence ${\bf g} = 1$, giving a sub-leading dependence $\sim N^{-2}$ because  of the higher genus.

The  connected four-point function (\ref{Gcconnec}) containing composite operators has ${\bf s}_c = m_1 + m_2$, but with $q = 6$ ramification points. The Riemann-Hurwitz formula (\ref{RiemHurwForm}) gives 
\be
{\bf g}_c = 0 ,
\ee 
but Eq.(\ref{largNCg}) shows that these functions also contribute at order $N^{-2}$ --- not because of a higher genus, but because of the higher number of ramification points.

\subsection{Covering maps.}	\label{SectCovMaps}


The use of the covering surface $\Sigma$ as a powerful tool for the computation of twisted correlation functions was introduced by Lunin and Mathur \cite{Lunin:2001pw}.
A covering surface $\Sigma$ of the base sphere $S^2_{\rm{base}}$, where $G(u,\bar u)$ is defined, is given by a map $z(t)$, with $t \in \Sigma$ and $z \in S^2_{\rm{base}}$, and with multiple inverses $t_{\frak a}(z)$ corresponding to the branches introduced by the twist operators in $G(u,\bar u)$. The ramification points ``replace'' the twist operators, so  on $\Sigma$, where there is only one single untwisted copy of the fields $X^i(t)$, $X^{i\dagger}(t)$, $\phi^i(t)$.

Here we want to find the genus-zero covering surface $\Sigma_c = S^2_{\rm{cover}}$ for the connected function $G_c$ in (\ref{Gcconnec}). We must explicitly construct a covering map $z :  S^2_{\rm{cover}} \to S^2_{\rm{base}}$ such that 
\bsub\label{modromicondsm1m2}
\begin{align}
&& z(t) &\approx b_1 t^{m_1}(t - t_0)^{m_2} &&  \text{as $z \to 0$}	\label{mondrlimtz0}
\\
&& z(t) &\approx 1+ b_2 (t - t_1)^{2} &&  \text{as $z \to 1$} \label{mondrlimtz1}
\\
&& z(t) &\approx u+ b_3 (t - x)^{2} &&  \text{as $z \to u$}	\label{mondrlimtzu}
\\
&& z(t) &\approx b_4 t^{m_1}  &&  \text{as $z \to \infty$}	\label{mondrlimtzinf}
\end{align}
\esub
The powers impose the correct monodromies of the inverse maps $t_{\frak a}(z)$ around the position of the twists in $z = \{0,1,u,\infty \} \in S^2_{\rm{base}}$. 
Because of the branching points, $\Sigma_c$ will have a number of sheets equal to the number of distinct elements entering the permutations in twists,  given by (\ref{scoloc}),
\be
{\bf s}_c = m_1 + m_2 . 	\label{numbofshee3}
\ee
It is a theorem in the theory of Riemann surfaces that a holomorphic map from the Riemann sphere to the Riemann sphere has the form $z(t) = f_1(t) / f_2(t)$,
where $f_1$ and $f_2$ are polynomials of degrees $d_1, d_2 \in {\mathbb N}$. 
From condition (\ref{mondrlimtzinf}), we know that 
$d_1 - d_2 = m_1$, hence
$d_1 > d_2$.
On the other hand, the larger degree $d_1$ is equal to the number of inverse maps $t_{\frak a}(z)$, hence to ${\bf s}_c$, so $d_1 = m_1 + m_2$.
To be consistent with (\ref{mondrlimtz0}), we thus must have $f_1 = A t^{m_1}(t - t_0)^{m_2}$. Also   $d_2 = m_1 - {\bf s}_c = m_2$, so $f_2 = B (t - t_\infty)^{m_2}$. Adjusting the constants $A$ and $B$ so that, as required by (\ref{mondrlimtz1}), $z(t_1) = 1$, we thus have
\be\label{coverm1m2}
z(t)=\left({t\over t_1}\right)^{m_1} \left( \frac{t-t_0}{t_1-t_0} \right)^{m_2} \left( \frac{t_1-t_\infty }{t-t_\infty} \right)^{m_2} .
\ee
%
Imposing that the map (\ref{coverm1m2}) locally satisfies the conditions (\ref{mondrlimtz1}) and (\ref{mondrlimtzu}) near the points $t_1$ and $x$ implies that 
\be
\begin{split}
\frac{1}{z} \frac{d z}{d t} &= {m_1 t^2+ [ (m_2-m_1)t_0-(m_1+m_2)t_\infty]t+m_1 t_0 t_\infty\over t(t-t_0)(t-t_\infty)} 
	= 0 
	\label{derz}
\end{split}	\ee
where the second equality holds at $z= t_1,x$.
In other words, $x$ and $t_1$ are the roots of the quadratic equation in the numerator. 
Using the relation between the coefficients and of this equation and its two roots, we find two relations between the parameters $t_1, t_0, t_\infty$ and $x$. We have the choice of fixing one of the $t_1, t_0, t_\infty$, and the two relations fix the other two as a function of $x$, which is the image of the ``free'' point $u$.
We choose
\be
\begin{split}	\label{t1t0tinf}
t_0 &= x-1 ,
\\
t_1 &= \frac{(x-1) (m_1+ m_2 x- m_2)}{m_1 + m_2 x} ,
\\
t_\infty &= x-{m_2 x\over m_2 x+m_1}
\end{split}\ee
leading to the map $u(x) = z(x)$
\be
u(x) = \Bigg( \frac{x+ \frac{m_1}{m_2}}{x-1} \Bigg)^{m_1+m_2} \Bigg( \frac{x}{x - 1 + \frac{m_1}{m_2} } \Bigg)^{m_1-m_2}  
.\label{ux}
\ee
The form of a ratio of polynomials is analogous to the map found by Arutyunov and Frolov in \cite{Arutyunov:1997gt}.
When $m_1 = m_2$, the map degenerates to a considerably simpler function
\be
u(x) = \left( \frac{x+ 1}{x-1} \right)^{2m}  
\qquad
(m_1 = m_2 = m) .
\label{uxm2m2m}
\ee

There is an evident asymmetry in the maps (\ref{coverm1m2}) and (\ref{ux}) when we exchange $m_1$ and $m_2$. This is because in our derivation of $z(t)$ it was convenient to place ramification points at $t = 0$ and $t = \infty$, and we chose to place the points of order $m_1$ at these locations. Of course, we could just as well have chosen to place the points of order $m_2$ at $t = 0, \infty$, in which case we would find 
\begin{align}
\tilde z(t) &= \left({t\over \tilde t_1}\right)^{m_2} \left( \frac{t-\tilde t_0}{\tilde t_1 - \tilde t_0} \right)^{m_1} \left( \frac{\tilde t_1 - \tilde t_\infty }{t - \tilde t_\infty} \right)^{m_1} ,	\label{altzt}
\\
\tilde u(x) &= \Bigg( \frac{x+ \frac{m_2}{m_1}}{x-1} \Bigg)^{m_2+m_1} \Bigg( \frac{x}{x - 1 + \frac{m_2}{m_1} } \Bigg)^{m_2-m_1}  .	\label{altux}
\end{align}
The maps $z(t)$ and $\tilde z(t)$ are isomorphic, one can pass from one to another with a M\"obius transformation, and describe the same covering surface $\Sigma_c$; this is shown in Appendix \ref{AppIsoMaps}. We are going to use henceforth the map (\ref{coverm1m2}). For must purposes we can assume, without loss of generality, that $m_1$ is the greater of $\{m_1,m_2\}$. 
Note that when $m_2=1$, corresponding to the trivial twist $\s_{(1)}$, our maps $z(t)$ and $u(t)$ reduce to the well-known expressions for non-composite operators (see e.g.~\cite{Lima:2020boh}).

One way of confirming the correctness of our covering map is to use the fact%
	\footnote{See e.g.~\cite{Pakman:2009zz}.}
	 that the number ${\bf H}_c$  --- known as the `Hurwtiz number' --- counting the different coverings of the sphere $S^2_{\rm{base}}$, with fixed number of ramification points of a fixed order, is equal to number of equivalence classes of permutations satisfying Eq.(\ref{mm2mmcond}) and the conditions for connectedness which lead to (\ref{scoloc}).
The number of different covering surfaces is equal to the number of solutions $x_{\frak{a}}(u_*)$, $\frak a = \frak{1}, \cdots, {\bf H}_c$ of the equation $u(x) = u_*$ for a fixed $u_*$.  
Inspection of the map (\ref{ux}) (or of the map (\ref{altux}) as well) shows that $u(x) = u_*$ reduces to a polynomial equation of order $2 \max(m_1,m_2)$, hence 
\be
{\bf H}_c = 2 \max(m_1,m_2).	\label{HurNumux}
\ee
The fact that, indeed, ${\bf H}_c$ is also the number of solutions to Eq.(\ref{mm2mmcond}) modulo global $S_N$ transformations is shown in Appendix \ref{AppHurNum}. Note that
${\bf H}_c$ is therefore the number of terms in the sum (\ref{GsumGen}). 



\subsection{Computation of the connected four-point function.}	\label{SectStressTensor}

We now use the covering maps to compute $G_c(u,\bar u)$,
following the `stress tensor method' 
\cite{Dixon:1986qv,Arutyunov:1997gt,Pakman:2009ab}. 
The Ward identity for the stress-energy tensor gives 
\be\label{method}
\begin{split}
F  &= \frac{\big\langle T(z)  R_{m_1}^-R_{m_2}^-  (\infty)\Oint (1)\Oint(u)  R_{m_2}^+R_{m_1}^+ ) (0) \big\rangle}{ \big \langle  R_{m_1}^-R_{m_2}^-  (\infty)\Oint(1) \Oint(u)  R_{m_2}^+R_{m_1}^+  (0) \big\rangle}
\\
	 &= \frac{1}{(z-u)^2} + \frac{H(u)}{z-u}+ \cdots	 
\end{split}
\ee
If one is able to obtain independently the function $H(u)$, then (\ref{method}) leads to a simple differential equation, 
\be\label{equ}
\p_u \log G(u) = H(u) ,
\ee
which determines the holomorphic part of $G(u, \bar u) = G(u) \bar G(\bar u)$;
 the anti-holomorphic part $\bar G(\bar u)$ is found by the analogous procedure with the anti-holomorphic stress-tensor $\tilde T(\bar z)$. 
The function $H(u)$ inherits the monodromy conditions of its twists, and is rather complicated. Nevertheless, with the aid of the covering map, one can find a function $H(x)$, parameterized by $x$, and solve the equation 
\be	\label{equx}
\pa_x \log G(x) 
			= u'(x) H(x) ,
\ee
obtained by a change of variables from $u$ to $x$ in (\ref{equ}).
To obtain $G(u,\bar u)$, we must then invert the map (\ref{ux}), 
$G(u,\bar u ) = C_{\bf 0} \sum_{\frak a} G(x_{\frak a}(u)) \bar G(\bar x_{\frak a}(\bar u))$.
The ${\bf H}_c$ inverses of $u(x)$ each correspond to a representative of one of the conjugacy classes in (\ref{GsumGen}).%
	\footnote{%
	The number ${\bf H}_c$ of equivalence classes is encoded in the covering map, as discussed at the end of \S\ref{SectCovMaps}, but the symmetry factors carrying the $N$-dependence in (\ref{GsumGen}) is not.}
The inverses $x_{\frak a}(u)$ can only be obtained locally, but for our purposes $G(x)$ is sufficient, and this can be found exactly. 
Let us show how.

We compute the equivalent of (\ref{method}) on the covering surface, namely
\be\label{methodcov}
\begin{split}
& F_{\rm{cover}} (t,x) =
	 \frac{\big\langle T(t) R^-(\infty)R^-(t_\infty)   \Oi(t_1,\bar t_1) \Oi (x,\bar x ) R^+(t_0) R^+(0) \big\rangle}{\big\langle R^-(\infty)R^-(t_\infty)   \Oi(t_1,\bar t_1) \Oi (x ,\bar x) R^+(t_0) R^+(0)  \big\rangle} .
\end{split}
\ee
Note how each part of the composite operator $R^+_{m_1} R^+_{m_2}(0)$ has been lifted to a different point on $S^2_{\rm{cover}}$, viz.~$R^+_{m_1}(0)$ goes to $t = 0$ and $R^+_{m_2}(0)$ goes to $t = t_0$, with a similar thing happening with $R^+_{m_1} R^+_{m_2}(\infty)$ being lifted to $\infty$ and $t_\infty$.
The absence of indices $m_1,m_2,2$ in (\ref{methodcov}) is because the twists are trivialized on $S^2_{\rm{cover}}$, $\s_{(n)} \mapsto 1$, and also $\sum_I \phi^a_{I} \mapsto n \phi^a$. Thus, for example, from (\ref{micro1}), we have
\be
R^{\pm}(t) = \exp \left( \pm \tfrac{i}{2}\big[ \phi^1(t) + \phi^2(t)  \big] \right) .
\ee

The pre-image $\Oi(t,\bar t)$ of the interaction operator (\ref{interaction}) is a sum of terms containing $\pa X^i(t)$ or $\pa X^{i\dagger}(t)$ and exponentials of $\phi^a(t)$, which can be expressed schematically as $\Oi (t) =  V_- + V_+$ where 
\be
\begin{split}
V_\pm(t, \bar t) &\equiv \Big[  ( \cdots ) \pa X (t) + c.c. \Big] \lor e^{\pm \frac{i}{2} \left[ \phi^1(t) - \phi^2 (t) \right]} \ror
\end{split}\label{OiVpm}\ee
with $(\cdots)$ containing anti-holomorphic fields including $\bar \pa X(\bar t)$ and exponentials of $\tilde \phi^a (\bar t)$. 
The complete expressions can be found e.g.~in \S2.3 of \cite{Burrington:2012yq}, but all we need here is the holomorphic fermionic factor $\lor e^{\pm \frac{i}{2} ( \phi^1 -  \phi^2 )} \ror$, and the fact that holomorphic%
	\footnote{%
	Of course, the same is true for the anti-holomorphic sector: $V_\pm(t,\bar t)$ can be organized instead as
	$
	V_\pm(t, \bar t) = \left[  ( \cdots ) \bar \pa X (\bar t) + c.c. \right] \lor e^{\pm \frac{i}{2} \left[ \tilde \phi^1(\bar t) -  \tilde \phi^2 (\bar t) \right]} \ror
	$, with holomorphic fields hidden in the ellipsis.
	}
 bosons always appear ``linearly'' as $\pa X$ or $\pa X^\dagger$. 
 This is sufficient for seeing that, after computing contractions, one can always rewrite expressions in the numerator of (\ref{methodcov}) as proportional to the correlation function in the denominator. See e.g.~\cite{Lima:2020boh,big_MAG}. 
 The final result is that
 \be
\begin{split}
F_{\rm{cover}} (t,x) &= \frac{(t_1 - x)^2}{( t - t_1 )^2 (t - x)^2} 
\\
	& + \frac{1}{4} \Bigg[ \left(\frac{1}{t - t_\infty} - \frac{1}{t-t_0} - \frac{1}{t} \right)^2
	+ \left( \frac{1}{t-t_1} - \frac{1}{t -x} \right)^2 \Bigg].
\end{split}		\label{FcoverRamond}
\ee

We now must map from $t$ to $z$ by inverting (\ref{coverm1m2}). 
Similar calculations can be found in \cite{Arutyunov:1997gt,Pakman:2009mi}; here we outline the main steps for the case of our map (\ref{coverm1m2}).
Taking the logarithm of the ratio $z(t) / z(x)$, we find the power series
\be
\begin{split}
& \sum_{k = 1}^\infty b_k (z - u)^k = (t-x)^2 \sum_{k = 0}^\infty a_k (t -x)^k 	
\\
& \text{hence}
\quad
t - x = \sum_{k =1}^\infty c_k (z - u)^{k/2} , 
\end{split}
\label{txSerzu}
\ee
where the $c_k$ can be solved order by order in terms of the coefficients $a_k$ and $b_k$. 
To find the pole in (\ref{method}), we just need the first three $c_k$, namely
\begin{align}
\begin{split}
c_1 = \mp \sqrt{\frac{b_1}{a_0}} ,
\quad
c_2 = - \frac{a_1 b_1}{2a_0^2} ,
\quad
c_3 = \mp \frac{ 5 a_1^2 b_1^2 - 4 a_0 a_2 b_1^2 +  4a_0^3 b_2}{8 a_0^{7/2} b_1^{1/2}}
\end{split}\label{ck123}
\end{align}
in which we must insert 
\begin{align*}
a_0 &= \frac{m_1 [m_1+m_2 (2 x-1)] }{2 m_2 x^2}
\\
a_1 &= -\frac{m_1 \left[m_1^2+3 m_1 m_2 x+m_2^2 \left(3 x^2 -1\right)\right]}{3 m_2^2 x^3}
\\
a_2 &= \frac{1}{4} \left(\frac{(m_1+m_2 x)^4}{m_2^3 x^4} -m_2-\frac{m_1}{x^4} \right)
\\
b_1&= \left(\frac{m_1+m_2 (x-1)}{m_2 x}\right)^{m_1-m_2} \left(\frac{\frac{m_1}{m_2}+x}{x-1}\right)^{-m_1-m_2}
\\
b_2 &= -\frac{1}{2} \left(\frac{m_1+m_2 (x-1)}{m_2 x}\right)^{2 (m_1-m_2)} \left(\frac{\frac{m_1}{m_2}+x}{x-1}\right)^{-2 (m_1+m_2)}
\end{align*}
We thus obtain two sets (the covering surface near $z = u$ has two sheets) of solutions $c_k(x)$, $k=1,2,3$.

The transformation of (\ref{FcoverRamond}) is governed by the transformation of the stress-tensor, \be
\begin{split}
F^{++} (z,x) 
	=  2 \big\{ t, z \big \} 
&	
	+ \left(\frac{dt}{dz}\right)^2  \frac{2(t_1 - x)^2}{\left( t(z) - t_1 \right)^2 \left( t(z) - x \right)^2} 
\\
&	+ \frac{1}{2} \left(\frac{dt}{dz}\right)^2   \Bigg[ \left(\frac{1}{t(z) - t_\infty} - \frac{1}{t(z) -t_0} - \frac{1}{t(z)} \right)^2
\\
&\qquad\qquad\qquad
	 + \left( \frac{1}{t(z)-t_1} - \frac{1}{t(z) -x} \right)^2 \Bigg]
\end{split}		\label{FzusumovaT}
\ee
where $\left\{t,z\right\}$ is the Schwarzian derivative, $\left\{t,z\right\}= ({t''\over t'})'-{1\over 2}({t''\over t'})^2$ and $t(z)$ is any of the two inverse maps  near $z = u$, given by (\ref{txSerzu}) with the two sets of solutions for $c_1,c_2,c_3$ --- both solutions give the same result, and their addition results in the factor of 2 appearing in the r.h.s.~of (\ref{FzusumovaT}).
Extracting the coefficient of the pole $\sim (z-u)^{-1}$ to get $H(x)$, and multiplying by $u'(x)$, we find the r.h.s.~of Eq.(\ref{equx}) as a function of $x$. Then, integrating Eq.(\ref{equx}), we obtain the connected four-point function
\be
\begin{split}
G_c^{++}(x) &=
	C_c^{++}   
	x^{1+ m_2-m_1}
	(x-1)^{2+m_1+m_2} 
\\	
&
	\times
	( x+ \tfrac{m_1}{m_2})^{2-m_1-m_2} 
	(x+ \tfrac{m_1-m_2}{ m_2})^{1+m_1-m_2} 
\\	
&
	\times
	(x+ \tfrac{m_1-m_2}{ 2m_2})^{-4}  .
\end{split}\label{ppfunc}
\ee

We have introduced indices $++$ in (\ref{ppfunc}) because we now want to distinguish the case for the other possible composite Ramond field, $R_{m_1}^+R_{m_2}^-$. 
The second-order correction of the two-point function of this neutral field is given by the same expression (\ref{4-point-R-1}) where now $G(u)$ has the form
\be
G^{-+}(u) = \big \langle R_{m_1}^-R_{m_2}^+(\infty)O_2^{int}(1)O_2^{int}(u)R_{m_1}^+R_{m_2}^-(0) \big \rangle.
	\label{gupm}
\ee
Instead of Eq.(\ref{FzusumovaT}), we now have
\begin{equation*}
\begin{split}
F^{-+} (z,u) =  2 \big\{ t, z \big \} 
&
	+ \left(\frac{dt}{dz}\right)^2  \frac{2(t_1 - x)^2}{\left( t(z) - t_1 \right)^2 \left( t(z) - x \right)^2} 
\\
&	+ \frac{1}{2} \left(\frac{dt}{dz}\right)^2  
	 \Bigg[ 
	 \left(
	 - \frac{1}{t(z) - t_\infty}
	 + \frac{1}{t(z) -t_0} 
	 - \frac{1}{t(z)} 
	 	\right)^2
\\
&\qquad\qquad\qquad
 + 
	 \left( 
	 \frac{1}{t(z)-t_1} 
	 - \frac{1}{t(z) -x} 
	 \right)^2 
	 \Bigg] ,
\end{split}		\label{FzusumovaTmp}
\end{equation*}
which leads to a different $H(x)$, and to the solution of (\ref{equx}) being
\begin{align}
\begin{split}
G^{-+}_{c}(x) &=
	C_c^{-+}  
	x^{2+m_2-m_1}
	(x-1)^{1+m_1+m_2} 
\\
&
	\times	
	( x+ \tfrac{m_1}{m_2})^{1-m_1-m_2} 
	(x+ \tfrac{m_1-m_2}{m_2})^{2+m_1-m_2}
\\
&
	\times	
	(x+ \tfrac{m_1-m_2}{2m_2})^{-4}  .
\end{split}		\label{pmfunc} 	\end{align}
Again, there are other contributions $G^{-+}_{m_1}$ and $G^{-+}_{m_2}$, coming from factorizations like in (\ref{G1factorzd})-(\ref{G2factorzd}). These non-composite four-point functions again reduce to what has been computed in \cite{Lima:2020boh}.

\subsection{Non-composite contributions and the full function.}	\label{SectNonCompositezuG}

One can use the stress-tensor method allied with the covering surface technique to compute the non-composite functions $G_{m_p}(u,\bar u)$, $p = 1,2$, as well; see \cite{Lima:2020boh,big_MAG}. 
As mentioned before, the covering surfaces of these functions have only four ramification points, and at genus zero the covering map is
\begin{align}
z_p(t) &= \left({t\over t_1}\right)^{m_p} \left( \frac{t-t_0}{t_1-t_0} \right) \left( \frac{t_1-t_\infty }{t-t_\infty} \right), 	\label{coverp}
\\
u_p(x) &= {x^{m_p-1}(x+m_p)^{m_p+1}\over (x-1)^{m_p+1}(x+m_p-1)^{m_p-1}} 
	\label{uxp}
\end{align}
where 
$t_0 = x -1$, $t_\infty = x - x (x+m_p)^{-1}$ and
$t_1 = t_0 t_\infty  / x$.
As mentioned, these maps can be obtained by making $m_1 = m_p$ and $m_2 = 1$ in (\ref{coverm1m2}) and (\ref{ux}), as it was to be expected.
The function $u_p(x)$ has ${\bf H}_p = 2m_p$ inverses.

Proceeding with the stress-tensor method, we find for the correlators (\ref{G1factorzd})-(\ref{G2factorzd}) 
\be\begin{split}
G_{m_p}(x) &= 
	C_{m_p} 
	x^{\frac{5(2-m_p)}{4}} 
	(x-1)^{\frac{5(2+m_p)}{4}}
	(x+m_p)^{\frac{2-3m_p}{4}}
\\
&
\times
	(x+m_p-1)^{\frac{2+3m_p}{4}}
	(x+\tfrac{m_p-1}{2})^{-4} .
\end{split} \label{old-R-func-p}	\ee

Restoring the symmetry factors $C_{\bf g}(N)$ given by Eq.(\ref{GsumGen}) in the genus expansion, and taking only the terms with ${\bf g} = 0$, the function $G(u,\bar u)$ in Eq.(\ref{Guu}) can be found from the functions $G(x)$ computed above as
\begin{align}
G(u,\bar u) &= 	
	\frac{\varpi(m_1)}{N}
	\sum_{\frak a = \frak1}^{2 m_1} G_{m_1}(x_{1,\frak a}(u)) \bar G_{m_1}(\bar x_{1,\frak a}(\bar u)) 
\nonumber \\
	&\quad
	+
	\frac{\varpi(m_2)}{N}
	\sum_{\frak a = \frak1}^{2 m_2} G_{m_2}(x_{2,\frak a}(u)) \bar G_{m_2}(\bar x_{2,\frak a}(\bar u)) 
\nonumber \\
	&\quad
	+
	\frac{\varpi(m_1 m_2)}{N^2}
	\sum_{\frak a = \frak1}^{2 \max(m_1,m_2)} G_c(x_{\frak a}(u)) \bar G_c(\bar x_{\frak a}(\bar u)) 
\label{Guufromx}
\end{align}
where $x_{p,\frak a}(u)$ are the inverses of $u_p(x)$.
The $N$-independent factors $\varpi$ will not be relevant for our discussion.
Note that, since (\ref{old-R-func-p}) is computed with the genus-zero map (\ref{coverp}), it only takes into account the terms of types \ref{Factr1} and \ref{Factr2b} discussed in \S\ref{SectConnDisconnFunc}. These are the leading contributions, at order $N^{-1}$, to the disconnected non-composite four-point functions. The terms of type \ref{Factr2a}, which contribute at order $~N^{-2}$ to (\ref{G1factorzd})-(\ref{G2factorzd}), must be computed with a map corresponding to the appropriately ramified genus-one covering surface.
In summary, Eq.(\ref{Guufromx}) contains every genus-zero contribution, but not every $N^{-2}$ contribution.

\section{OPEs and fusion rules}	\label{SectOpeStrConstants}


We now use Eq.(\ref{Guufromx})
to examine various possible OPEs, by taking the coincidence limit of the operators in the four-point functions. 
Expressing the functions $G_c(x)$ as explicit functions of $u$ is impossible in general, because  one would need to know all the inverses $x_{\frak a}(u)$, but to find the OPEs we only need to invert the functions locally, which can done by expanding  the functions $u(x)$ near the singular points. 
Exploring the OPE channels gives us one more check of formulae (\ref{ppfunc}) and (\ref{pmfunc}) for $G^{\pm +}_c(x)$, since they must yield consistent fusion rules with the known ones for the disconnected functions. Also, the OPEs allow us to fix the integration constants $C^{\pm +}_c$ and $C_{m_p}$ which are undetermined by the stress-tensor method.


We first consider the contributions from the  connected functions $G^{++}_c(x)$ and $G^{-+}_c(x)$ given in Eqs.(\ref{ppfunc}) and (\ref{pmfunc}).
For each function, we analyze two short-distance behaviors: the limit $u \to 1$ corresponding to the OPE $\Oint(u)\Oint(1)$, and the limit $u \to 0$, corresponding to the OPE between $\Oint(u)$ and the composite Ramond operator.


\subsection{OPE of two deformation operators.}

Let us start with the OPE of two interaction terms $\Oint(u)\Oint(1)$. This corresponds to taking the limit $u\to 1$ in the correlation function (\ref{Gcconnec}). 
Among the solutions of $u(x) = 1$, only two contribute non-singular terms in the expansion of $G_c(x)$, namely  $x\to \infty$ or $x\to {m_2-m_1\over 2m_2}$ (they correspond to $x\to t_1$). 
Each of these solutions give a different OPE channel, corresponding to a different conformal family in  $[\Oint] \times [\Oint]$.
Let us consider first $x\to \infty$; inverting $u(x)$ asymptotically, it follows that 
\be
x(u) = -{4m_1\over 1-u}+{1\over 2} \left( 1+4m_1-{m_1\over m_2} \right)+ \rm{O}(1-u)	.\label{xnearinfu1}
\ee
Expanding $G^{++}_c(x)$ accordingly, we get 
\begin{align}
G^{++}_c(x(u)) &= C_c^{++} x^2 \left[ 1- \left( 1+4m_1- \tfrac{m_1}{m_2} \right) \frac{1}{x} + \rm{O}(\tfrac{1}{x^2}) \right] \nonumber
\\
	&= \frac{16 m_1^2 C_c^{++}  }{ (1-u)^2} + 0\times {1\over 1-u}+ \text{non-sing.} \label{xinf}
\end{align}
From counting the dimensions, it is clear that this channel corresponds to the identity operator, i.e. $[\Oint ] \times [ \Oint(1) ] \sim [ \mathds 1 ]+ \cdots$. The absence of  sub-leading singularities ensures that there is no  operator of dimension $1$ in this OPE, as it should be for a truly marginal deformation.
Taking each ``component'' $\Oi_{(k \ell)}$ of the $S_N$-invariant operator $\Oint$ to have a normalized two-point function, 
\be
\big\langle \Oi_{(k \ell)}(1) \Oi_{(k \ell)}(u,\bar u) \big\rangle = \frac{1}{|1-u|^4} ,
\ee
which can always be done by adjusting the deformation parameter $\la$, the structure constant of this OPE is one, and inserting it back into the four-point function we find that 
\be\begin{split}
\frac{1}{(1-u)^2} 
&= \frac{\big\langle R^-_{(m_1)} R^-_{(m_2)}(\infty) R^+_{(m_2)} R^+_{(m_1)}(0) \big\rangle}{(1-u)^2} 
\\
&=  \frac{16m_1^2 C_c^{++}}{(1-u)^2} ,
\end{split}\ee
where the two-point function is inherited from the original four-point function representative of the conjugacy class in this channel. Hence
\be
C^{++}_c = \frac{1}{16m_1^2} .		\label{Cppm1}
\ee

Now let us consider the terms that appear in the second channel. Inverting $u(x)$ near $x = {m_2-m_1\over 2m_2}$,
\begin{align}
x(u) -  {m_2-m_1\over 2m_2}=\left( {3\over 64}{(m_1^2-m_2^2)^2\over m_1m_2^4} \right)^{{1\over 3}}(1-u)^{{1\over 3}}+ \cdots	\label{xm2m2m1a}
\end{align}
Expanding $G_c^{++}(x)$ around $x = {m_2-m_1\over 2m_2}$,  we get to the following behavior of the function in this channel,
\be\label{chan23}\begin{split}
G^{++}_c(x(u)) &=  {C \over (1-u)^{4/ 3}} + 0\times {1\over 1-u}
	+ \frac{b}{(1-u)^{2/3}} + \frac{a}{(1-u)^{1/3}} + \text{non-sing.}
\end{split}\ee
where $a,b,C$ are constants, with
\be \begin{split}
C 
	&= \frac{4 C_c^{++} }{3^{\frac{4}{3}}} \frac{(m_1 + m_2)^2}{m_2^2} \left( \frac{m_1^4 m_2^4}{(m_1^2 - m_2^2)^2} \right)^{\frac{1}{3}} 
\\
&
	= \frac{1}{4 \cdot 3^{\frac{4}{3}}} \frac{(m_1 + m_2)^2}{m_2^2 m_1^2} \left( \frac{m_1^4 m_2^4}{(m_1^2 - m_2^2)^2} \right)^{\frac{1}{3}} .
\end{split}	\ee
Note that once we come back to the base sphere, the asymmetry in $m_1, m_2$ introduced by our choice of covering map disappears (after taking into account Eq.(\ref{Cppm1})).
Dimensional analysis of the leading term $\sim(1-u)^{4/3}$ in (\ref{chan23}) determines that this channel corresponds to the OPE
$\Oint \Oint \sim C_{223} \, \sigma_3 + \cdots$. The appearance of the twist field $\sigma_3$ is not surprising because the interaction $\Oint$ is constructed using $\sigma_2$, and the above OPEs follow the $S_N$ group multiplication rule $\sigma_2\sigma_2\sim \mathds 1+\sigma_3$. The sub-leading term in (\ref{chan23}) would correspond to an operator of dimension one, and its absence is again a confirmation of the correct behavior of the function $G^{++}_c(u,\bar u)$.

\bigskip

The behavior in Eqs.(\ref{xinf}) and (\ref{chan23}) matches precisely the one found for the disconnected functions $G_{m_p}$,  described in detail in \cite{big_MAG}.
Such consistency of the fusion rule 
$$
[\Oint] \times [\Oint] = [\mathds 1] + [\s_3]
$$
is another check of the  connected function (\ref{ppfunc}).
For the disconnected functions, the identity channel gives that
\be
C_{m_p} = \frac{1}{16 m_p^2} 
\ee 
by the same argument as above.

\subsection{OPE of the deformation operator and the composite Ramond field.}

Let us turn to the limit $u\to 0$. It corresponds to the OPE of the interaction field with the composite Ramond field: $\Oint (u)R_{m_1}^+R_{m_2}^+(0)$.
 Solving $u(x) = 0$, we find the channels
\bsub\begin{align}
x &\to -{m_1/ m_2} , \quad && \text{for} \quad m_1 \lessgtr m_2 \label{chnbth}
\\
x &\to 0 && \text{for} \quad m_1 > m_2	\label{chna}
\\
x &\to  (m_2-m_1)/  m_2 && \text{for} \quad m_1 < m_2	\label{chnb}
\end{align}\esub
Let us consider the common channel (\ref{chnbth}) first,
$$
x(u) + {m_1/ m_2}=c_1 u^{1\over m_1+m_2}+c_2 u^{2\over m_1+m_2}+ \cdots
$$
where the coefficients $c_i$ are readily computable. From here one gets for the correlation function in this channel,
\be\label{ch+}
G^{++}_c(u)= m_1^{{m_2-3m_1\over m_1+m_2}} m_2^{{m_1-3m_2\over m_1+m_2}} \
 u^{-1+{2\over m_1+m_2}}+ \cdots
\ee
with $C$ (another) constant.
Dimensional analysis of (\ref{ch+}) shows that the OPE in question has the following possible forms:
$$
\Oint (u)R_{m_1}^+R_{m_2}^+(0) \sim X\sigma_{m_1+m_2}(0)
$$
where $X$ is some operator of dimension 
\be
\Delta_X={9/4 \over m_1+m_2}	\label{Xdim}
\ee
and R-charge 1, acing on the twist field, or
$$
\Oint (u)R_{m_1}^+R_{m_2}^+(0)\sim \tilde X R^+_{m_1+m_2}(0)
$$
where $\tilde X$ has dimension 
\be
\Delta_{\tilde X}={2\over m_1+m_2}	\label{tilXdim}
\ee
and R-charge $1/2$. This second form should be connected to  previous results \cite{Carson:2014yxa,Tormo:2018fnt} where similar three-point functions, but with the chiral field $O_2$, instead of its descendent $\Oint$, were considered. In both cases, the numerical factor in (\ref{ch+}) plays the role of (the square of) the structure constant.

In the channel (\ref{chna}), where $m_1 > m_2$, we have
$$
x=c_1 u^{{1\over m_1-m2}}+c_2 u^{{2\over m_1-m2}}+ \cdots
$$
leading to
\be\label{ch-}
G^{++}_c (u)=C u^{-1+{1\over m_1-m_2}}+ \cdots
\ee
Once again, we expect to find a twist $\s_{m_1+m_2}$ in this channel, since this is the only possible combination of the twists $\s_{m_1}$, $\s_{m_2}$ and $\s_{2}$ in the conjugacy classes that compose the  connected function --- i.e.~the twist $\s_{2}$ joins the other two cycles. The exponent above implies that we can therefore have the OPE
$$
\Oint(u)R_{m_1}^+R_{m_2}^+(0)\sim Y \sigma_{m_1+m_2}(0)
$$
where $Y$ is now some operator of dimension 
$
\Delta_Y	= \frac{ {5\over 4} m_1}{m_1^2 -m_2^2} +  \frac{ {3\over 4} m_2 }{ m_1^2 -m_2^2}
$
and R-charge 1. 
Alternatively, we could also find, as above,
$$
\Oint(u)R_{m_1}^+R_{m_2}^+(0)\sim \tilde Y R^+_{m_1+m_2}(0)
$$
where now $\tilde Y$ has dimension $\Delta_{\tilde Y}={1\over m_1-m_2}$, and R-charge $1/2$.

If we finally look to the third channel (\ref{chnb}), we find nothing new: we get simply the results for channel (\ref{chna}), but with $m_1$ and $m_2$ interchanged. This was to be expected as, we emphasize, the functions on the base sphere are symmetric under $m_1 \leftrightarrow m_2$.

\subsection{The case of $m_1= m_2$.}	\label{Sectm1eqm2}

Now let us consider  the behavior of our correlation function when $m_1=m_2=m$. 
The highly simplified $u(x)$ is now given by Eq.(\ref{uxm2m2m}).
We can compute the correlation function with the same procedure as before, and find simply
$$
G^{++}_c(x)=C_c^{++} x^{-2}(x-1)^{2m+2}(x+1)^{-2m+2}.
$$
In the limit $u\to 1$ with $x\to\infty$ we find again a behavior showing that the identity appears in the product of interaction fields, and in the other limit, $u\to 1$ with $x\to 0$, the coefficient in front of the contribution of the field $\sigma_3$ vanishes, so in this case there is no such channel in the OPE of two interaction terms.
When $u\to 0$, one single solution survives: $x\to -1$, and the function scales as
$$
G^{++}_c(u) = c \; u^{-1+{1\over m}}+ \cdots
$$
This means that, if one accepts our suggestions above, only  descendants  of $\sigma_{2m}$ or $R_{2m}$ appear on the r.h.s.~of the OPE, and the term like $\sigma_0$ is absent, as it should be, of course.

\subsection{OPEs from the four-point function with neutral composite operators.}

We turn next to consider  the short-distance behavior of the two-point function  (\ref{pmfunc}) of the neutral composite fields $R_{m_1}^+R_{m_2}^-$. 
Its behavior as $u\to 1$, corresponding to the OPE of the two interaction terms, is exactly the same as discussed above, as expected for consistency,
and yields $C_c^{+-} ={1/ 16 m_1^2}$ in the identity channel.

The limit $u\to 0$ accounts for the OPE  $\Oint(u)R_{m_1}^+R_{m_2}^-(0)$. In the channel $x\to -{m_1/ m_2}$,
$$
G^{+-}_c(u) \sim C  u^{-1+{1\over m_1+m_2}}+ \cdots
$$
This result leads to the following possible suggestions for the OPE:
$$
\Oint(u)R_{m_1}^+R_{m_2}^-(0)\sim \scr Y  \sigma_{m_1+m_2}(0)
$$
where $\scr Y$ is some operator of dimension $\Delta_{\scr Y} = {5/4\over m_1+m_2}$ and R-charge zero, or
$$
\Oint(u)R_{m_1}^+R_{m_2}^-(0)\sim \tilde {\scr Y} R^\pm_{m_1+m_2}(0)
$$
with $\tilde {\scr Y}$ having $\Delta_{\tilde {\scr Y}}={1\over m_1+m_2}$, and R-charge $\mp 1$.

The channel $x\to 0$ leads to 
$$
G^{+-}_c(u)\sim C u^{-1+{2\over m_1-m_2}}+ \cdots
$$
and one possible interpretation of this scaling for the form of the OPE is
$$
O_2^{int}(u)R_{m_1}^+R_{m_2}^-(0)\sim {\scr X} \sigma_{m_1+m_2}(0)
$$
where ${\scr X}$ has $\Delta_{X}={9/4\over m_1+m_2}$ and R-charge zero; alternatively,
$$
\Oint(u)R_{m_1}^+R_{m_2}^-(0)\sim \tilde {\scr X} R^\pm_{m_1+m_2}(0)
$$
with $\tilde {\scr X}$ having $\Delta_{\tilde {\scr X}}={2\over m_1+m_2}$, and R-charge $\mp 1$.
Notice that ${\scr X}$ has the same dimension as found for the operator $X$ in Eq.(\ref{Xdim}), but $X$ is R-charged while $\scr X$ is R-neutral. Similarly, $\tilde{\scr X}$ and $\tilde X$ have equal dimensions given by (\ref{tilXdim}), but different R-charges.




\section{Renormalization and anomalous dimensions}	\label{RenAnomDim}

 The two-point function of the composite Ramond fields  $R_{m_1}^{\pm}R_{m_2}^{\pm}$,  evaluated at second order in the deformed orbifold SCFT$_2$  (\ref{correct}), contains, in the large-$N$ limit, a $\log {|z_{14}|}$ correction term together with the logarithmic divergence
\be
\begin{split}
\lambda^2 \pi \log \Lambda \int \! d^2u \, G(u,\bar u)
&	= \lambda^2 \pi \log \Lambda 
	\Bigg[
	\int \! d^2x \left| u'(x) \, G_c(x) \right|^2 
\\
	&\quad
	+
	\int \! d^2x \left| u'_{1}(x) \, G_{m_1}(x) \right|^2  
	+ 
	\int \! d^2x  \left| u'_{2}(x) \, G_{m_2}(x) \right|^2 
	\Bigg] . 
\end{split}
\label{log-div}
\ee
%
%
For each of the three functions composing (\ref{Guu}), we have made a change of integration variables $d^2u=d^2x |u'(x)|^2$ with the maps $u(x)$  given in Eqs.(\ref{ux}) and (\ref{uxp}).
We are forced to do this change of variables, since we have calculated in (\ref{ppfunc}), (\ref{pmfunc}) and (\ref{old-R-func-p}) the explicit form  of the correlation functions parameterized by $x$. 

We start with the integral
\be
\begin{split}
I^{++}_{c} 
	&= \int \! d^2x  |u'(x)G_c^{++}(x)|^2 
\\
	&\sim \int d^2x  \left| {(x-1)(x+{m_1\over m_2})\over (x+{m_1-m_2\over 2m_2})^2} \right|^2
\end{split}\ee
and, with one more  change of variables,
$$
y=- 4m_2 (m_1+m_2)^{-2} (x-1)(m_2 x +m_1),
$$
we arrive at
\be
I^{++}_c \sim \int d^2y {|y|^2\over |1-y|^3} = \frac{1}{\Gamma(-1)} = 0.	\label{Ivanishpp}
\ee
The same happens in the case of  R-neutral composite Ramond field $R_{m_1}^+R_{m_2}^-$ --- 
now $G^{+-}_c(x)$ is given by Eq.(\ref{pmfunc}) and its integral is
\be
\begin{split}
I^{+-}_c &= \int \! d^2u \; G^{+-}_c (u,\bar u) 
\\
	&= \int \! d^2x |u'(x)G_{c}^{-+}(x)|^2 
\\
	&\sim \int d^2x \left| {x(x+{m_1-m_2\over m_2})\over (x+{m_1-m_2\over 2m_2})^2} \right|^2 .
\end{split}
\ee
Again by a further change of the variables, 
$$
y(x) =  - 4m_2  (m_1-m_2)^{-2} \left( x+{m_1-m_2\over m_2} \right)  x, 
$$
we get  exactly the same result  as before,
\be
I^{+-}_c \sim \int d^2y {|y|^2\over |1-y|^3} = 0.	\label{Ivanishpm}
\ee
Hence \emph{the  connected part  $G_c$  of the four-point function (\ref{4-point-R-1}) does not contribute to the anomalous dimensions of any of the considered composite operators.}


We next compute the contributions  coming from  the  last two terms, $G_{m_1}(x)$ and $G_{m_2}(x)$, in  Eq.(\ref{log-div}), i.e.~the disconnected part of the function. 
Using (\ref{old-R-func-p}) and (\ref{uxp}), the last two integrals in Eq.(\ref{log-div})  take the form  \cite{Lima:2020boh}
\be
\begin{split}
 J_R(n)  =  \left(\frac{n+1}{16 n}\right)^2  \int \! d^2y \;  |y|^{2a}|1-y|^{2b}|y-w_n|^{2c},  
\\
w_n \equiv \frac{4n}{(n+1)^2} ,
\end{split}
\ee
where $n = m_1$ or $n = m_2$,
and $a = \tfrac{1}{2} + \tfrac{1}{4}n$, $b = - \tfrac{3}{2}$,  $c = \tfrac{1}{2} - \tfrac{1}{4}n$.
Evaluation of the above integrals  $J_R(m_p)$ can be performed by applying   the Dotsenko-Fateev method \cite{Dotsenko:1984nm,Dotsenko:1984ad}.
 The final result can be written in terms of combinations of hypergeometric functions which asymptote to finite, small numbers when $n$ is large  \cite{Lima:2020boh}.

The first consequence of the existence of finite non-vanishing terms in Eq. (\ref{log-div}) is the renormalisation of the conformal dimensions of the composite twisted Ramond fields.
In order to cancel the $\log \Lambda$ divergent terms, we follow the standard QFT rules, i.e.~dressing each one of the ``bare'' Ramond fields to get their renormalized counterparts
\be
R^{\pm(ren)}_{m_p} (z,\bar z) = \Lambda^{\frac{1}{2} \pi \lambda^2 J_R(m_p)} R^{\pm}_{m_p} (z,\bar z).
\label{dressedR}
\ee 
Therefore the $\lambda^2$-corrected conformal dimensions of the composite Ramond fields in deformed orbifold SCFT$_2$ takes the form 
\be
\begin{split}
\Delta^R_{m_1,m_2} (\lambda) + \tilde \Delta^R_{m_1,m_2} (\lambda)
	&= \frac{m_1+m_2}{2} 
	+ \frac{1}{2} \pi \lambda^2 \big(   | J_R(m_1)| + |  J_R(m_2)| \big),
\label{anomal-dim}\end{split}
\ee
and the two-point functions  of the composite Ramond fields can be rewritten as
\begin{align}
\begin{split}
&\frac{\big\langle R_{m_1}^-R_{m_2}^-(z_1,\bar{z}_1) \, R_{m_1}^+R_{m_2}^+(z_4,\bar{z}_4) \big\rangle^{ren}_{\lambda}} {\big\langle  \mathds 1  \big\rangle _{\lambda}} 
\\
&
= \frac{1}{|z_{14}|^{m_1+m_2 + \pi \lambda^2( |J_R(m_1)| +|J_R(m_2)|)}} 
\\
&
= \frac{1}{|z_{14}|^{m_1+m_2} } \Big[ 1 - \pi \lambda^2 ( |J_R(m_1)| +|J_R(m_2)|) \log {|z_{14}|}
\\
&\qquad\qquad\qquad\qquad
 + O(\lambda^4) \Big] .
\end{split}
\label{2-point}
\end{align}
A similar renormalization occurs for the R-neutral composite Ramond fields $R_{m_1}^+R_{m_2}^-$; in fact  both type of composite Ramond fields (charged and  neutral)  turn out to have equal conformal dimensions, but different R-charges.

We have to note another important implication of the above result, concerning the non-vanishing finite parts in the  integral in Eq. (\ref{log-div}). It  allows one to also derive the  non-zero correction to the three-point function  
\be
\begin{split}
&\big\langle R_{m_1}^-R_{m_2}^-(\infty) \; \Oint(1) \; R_{m_1}^+ R_{m_2}^+(0) \big\rangle_{\lambda}=
  \lambda \Big( J_R(m_1) + J_R(m_2) \Big) + \cdots,
		\label{3-point-RR}
\end{split}
\ee
which in fact is providing the value  of  the structure constant at the first order in perturbation theory in $\la$.


The fact that at the second order in perturbation theory  the purely connected part $G_c(x)$  of the $S_N$ invariant 4-point function (\ref{4-point-R-1}) gives  no contributions  to the two-point function of the composite Ramond fileds $R_{m_1}^{\pm}R_{m_2}^{\pm}$, while those of the so-called ``disconnected'' parts $G_{m_p}(x)$  yield non-vanishing contributions raises the question: 
{\it Could one  impose appropriate restrictions on the values of the twists $m_p$ 
 that select the BPS-protected from the lifted (non-protected) composite  Ramond states?}

The answer is hidden in the structure of cycles entering connected and partially-disconnected functions, as described in \S\ref{SectConnDisconnFunc}. 
For an operator to be protected, it must only posses the  connected part of $G(u,\bar u)$, hence the cycles $(m_1)(m_2)$ must be such that the partially-disconnected functions are impossible.
We get a partially-disconnected function $G_{m_1}(x)$ when a cycle of the deformation operator, $(2)=(k \ell)$, with  $k, \ell \in [1,N]$, 
is such that one of its elements, say $\ell$,  coincides with one of the elements of the cycle $(m_1)$ and the second one, $k$, \emph{does not} belong neither to $(m_1)$ nor to $(m_2)$.  
Similarly, the function $G_{m_2}(x)$ is made of terms with  $\ell \in (m_2)$. It is then clear that, when the cycles $(m_1)$ and $(m_2)$ are such that 
\be
m_1+m_2=N,	\label{protect}
\ee
there is no $k\in [1,N]$ which does not belong to either cycle $(m_p)$, hence we have no disconnected contributions. Thus the family of composite fields  $R_{(m_1)}^{\pm} R_{(m_2)}^\pm$ satisfying (\ref{protect}) is \emph{protected}: they do not receive any corrections to their ``free orbifold point''  conformal dimensions 
$\Delta^R_{m_1,m_2} + \tilde \Delta^R_{m_1,m_2} =\frac{1}{2} N$. 
In all other cases, since $m_1+m_2 < N$, one is able to choose $k \in [1,N]$ that  is not in $(m_1)$ nor in $(m_2)$. Then we have both connected and partially-disconnected contributions to the four-point functions and, as a result  these composite Ramond states (and fields) are \emph{lifted}, i.e.~they get $\la^2$ dependent corrections (\ref{anomal-dim}) to their conformal dimensions.

Note that, while we have been considering the $S_N$-invariant operator $R^\pm_{m_1} R^\pm_{m_2}$, we could also ask the fate of ``individual'', non-$S_N$-invariant operators 
$R^\pm_{(m_1)} R^\pm_{(m_2)}$, made by individual cycles $(m_1)$ and $(m_2)$, with no sum over group orbits.
It is not hard to see that the discussion above still holds: only operators with cycles satisfying Eq.(\ref{protect}) are protected. All other individual operators $R^\pm_{(m_1)} R^\pm_{(m_2)}$ undergo a renormalization of their dimensions obeying Eq.(\ref{anomal-dim}).
This is because the deformation action, and hence $\Oint$, must necessarily be an $S_N$-invariant object, hence we must always sum over the group orbits of the cycles $(2) = (k \ell)$ of the deformation operator, and $(k \ell)$ will always assume all possible values. 



\section{Concluding Remarks} \label{SectConclRemarks}

Coherent superpositions of twisted Ramond states are an important ingredient in the holographic duality between the two-charge extremal black hole solutions of type IIB supergravity and  the VEVs of operators in the SCFT$_2$ \cite{Skenderis:2008qn}. Comparison between  the bulk SUGRA solutions and the D1-D5 orbifold SCFT$_2$ data  is based on the conjecture that every chiral NS field $O_n$ and certain BPS  twisted Ramond ground states  are not affected by the marginal interaction (\ref{def-cft}), i.e.~the values of such VEVs are $\lambda$-independent.

In the broadly used interpretation of twisted states in terms of multi-winding of $m_i$-component strings, the composite operators $\prod_{i=1}^q R^\pm_{m_i} (z,\bar z)$, with $\sum_{i=1}^q m_i = N$,
correspond to twisted Ramond ground states%
	\footnote{%
	In the notation of Ref.\cite{Giusto:2012yz}.}
$\ket{0^{\pm\pm}}_R$ of the  
orbifold 
SCFT$_2$ with central charge $c_{orb} = 6N$.
The double-cycle operators considered here --- the winding of only two components, i.e.~$q = 2$ and $m_1 + m_2 = N$ --- are the simplest example of such ground states, apart from the maximal-twist single-cycle Ramond fileds $R^{\pm}_N$  \cite{Lima:2020boh}.
The selection rule we have found means that, while there \emph{is} a renormalization of individual states $\ket{0_{m_i}^{\pm\pm}}_R$,
 corresponding to the $m_i$-component strings, the double-wound states $\ket{0^{\pm\pm}}_R$
 with total weight $\Delta = \frac14 N$, composed by two $m_i$-component strings with $m_1 + m_2 = N$, is protected.
 For now, the renormalization properties of products  of more than two operators is still an open question,  but our preliminary investigations suggest that the ``double-winding'' selection rule  generalizes to multi-wound states. This is, indeed, the behavior expected for $\ket{0^{\pm\pm}}_R$ composed by several component strings:
the non-lifting of multi-wound Ramond ground states with weight $\Delta = \frac14 N$ has been used to identify them with two-charge geometries.
Operators $\big[ R^\pm_{k} (z,\bar z) \big]^{N/k}$ are holographically dual to axially-symmetric bulk geometries with a ${\mathbb Z}_{k}$ orbifold singularity at the end of the long AdS$_3\times S^3$ throat; hence the operators with $m_1 = m_2 = m = \frac{1}{2} N$, described in the present paper, yield a geometry with a conic singularity of ${\mathbb Z}_{N/2}$ type, see \cite{Giusto:2012yz}.
The $\ket{0^{\pm\pm}}_R$ are also related to excited states of the D1-D5-P system via appropriate integer or fractional spectral flows,  respectively describing ``neck'' or ``cap'' degrees of freedom in the three-charge geometries  \cite{Giusto:2012yz}.
Note that it is not hard to find the four-point functions for these excited states, given the ground state functions described here.

Let us mention a few more open problems that are under investigation. The first is the renormalization and the protection rules of R-neutral (but ``internal'' SU(2) doublets) twisted Ramond fields $R^0_n$, and of the corresponding R-neutral composite operators, such as $(R^0_n)^2$ and $R^{\pm}_m R^0_n$.
These fields, and their (left-right asymmetric) descendants, are important for the construction of  microstates of the three-charge extremal black hole in the D1-D5-P system \cite{Bena:2015bea}.
Another open question is about eventual $\la$-dependent 
changes to three-point functions which are  appropriate generalizations of  $\big\langle R_{m_1}^-R_{m_2}^-(\infty)   O_2 (1)  R_{m_1 +m_2}(0) \big\rangle_{\lambda}$, as for example those considered in the recent papers \cite{Tormo:2018fnt,Tormo:2019yus}.

To conclude, the problems solved in the present paper are based on the construction of the appropriate covering maps and the derivation of the renormalization of  two- and three-point functions involving  composite twisted Ramond fields in the deformed D1-D5 orbifold SCFT$_2$. An important byproduct of our investigations is a simple selection rule that allows us to separate between protected and lifted states. These results can be easily generalized for composite twist fields $\sigma_{m_1} \sigma_{m_2}$ and for  chiral NS fields $O_{m} O_{n}$, since the covering map to be used is the same as the one we have constructed (\ref{ux}). Our preliminary results indicate that the case of twist fields seems to be identical to the Ramond case, while composite  chiral NS fields, similarly to the single-cycle $O_n$ fields, seem to be free of any renormalization \cite{big_MAG}.

 In fact, the most important problem behind the question about the origins and the specific features of the protected and  lifted states is the lack of a complete description of the (super)symmetry algebra of the deformed orbifold SCFT$_2$, and the lack of knowledge of the structure of  its null vectors  and the eventual classification of its unitary representations. Many partial recent results \cite{Burrington:2018upk,deBeer:2019ioe,Dei:2019osr,Ahn:2020rev}  provide  important hints about different aspects of this problem. We believe that the information  extracted from the specific 3-, 4- and  5-point functions of (composite) twisted Ramond fields in the free orbifold point,  together with the developments of the methods of the calculations of certain integrals of them,  also  might  provide relevant indications about the spectra of the representations of the deformed D1-D5 orbifold model.

\bigskip

\no {\bf Acknowledgements}

\no 
We would like to thank an anonymous referee for insightful comments.
The work of M.S.~is partially supported by the Bulgarian NSF grant KP-06-H28/5 and that of M.S.~and G.S.~by the Bulgarian NSF grant KP-06-H38/11.
 M.S.~is grateful for the kind hospitality of the Federal University of Esp\'irito Santo, Vit\'oria, Brazil, where part of his work was done.

\appendix

\section{Isomorphism between covering maps}	\label{AppIsoMaps}

\setcounter{equation}{0}
 \renewcommand{\theequation}{\thesection.\arabic{equation}}

To proof that the maps $z(t)$ and $\tilde z(t)$ in (\ref{coverm1m2}) and (\ref{altzt}) are isomorphic, we must show that they are related by an automorphism of the cover $S^2_{\rm{cover}} = {\mathbb C} \cup \infty$. In other words, we must show that there is a M\"obius transformation $f : S^2_{\rm{cover}}  \to S^2_{\rm{cover}}$ such that 
\be
z = \tilde z \circ f .
\ee
Composing the M\"obius transformation
\be
f(t) = \frac{a t + b}{c t + d} \ , \qquad ad - bc \neq 0 \ ,
\ee
with the function $\tilde z$, given by (\ref{altzt}), we have
\begin{align*}
\tilde z \circ f(t) 
	&= \left( \frac{1}{\tilde t_1} \frac{a t + b}{c t + d} \right)^{m_2}
	 	\left( \frac{\frac{a t + b}{c t + d} -\tilde t_0}{\frac{a t + b}{c t + d} - \tilde t_\infty} \right)^{m_1} 
		\left( \frac{\tilde t_1 - \tilde t_\infty }{\tilde t_1 - \tilde t_0} \right)^{m_1} 
\\
	&= \left( \frac{1}{\tilde t_1} \frac{a t + b}{c t + d} \right)^{m_2}
	 	\left( \frac{ (a - \tilde t_0 c) t + b - \tilde t_0 d  }{ (a - \tilde t_\infty c) t + b - \tilde t_\infty d  } \right)^{m_1} 
			\left( \frac{\tilde t_1 - \tilde t_\infty }{\tilde t_1 - \tilde t_0} \right)^{m_1} 
\end{align*}
and we must find $a,b,c,d$ such that this equals
$$
z(t) = \left({t\over t_1}\right)^{m_1} \left( \frac{t-t_0}{t_1-t_0} \right)^{m_2} \left( \frac{t_1-t_\infty }{t-t_\infty} \right)^{m_2} .
$$
By inspection, the parameters must satisfy the conditions
\begin{align*}
a/c = \tilde t_\infty , 
\quad
b/d = \tilde t_0 ,
\quad
b/a = - t_0 ,
\quad
d/c = - t_\infty 
\end{align*}
hence 
$$
ad - bc =  \frac{t_\infty - t_0}{t_0} \, bc = \frac{bc}{(x-1)(1+m_2 x / m_1)} \neq 0 
$$
for $x \neq \infty$.

\section{Combinatorial derivation of the Hurwitz number for  connected functions}	\label{AppHurNum}

\setcounter{equation}{0}
 \renewcommand{\theequation}{\thesection.\arabic{equation}}

Here we show that 
\be
{\bf H}_c = 2 \max (m_1 , m_2)	\label{Hnumberm1m2}
\ee
%
by counting  how many different equivalence classes of permutations of the kind
\be
(m_1)_\infty (m_2)_\infty (2)_1 (2)_u (m_2)_0 (m_1)_0  = 1 ,	\label{PermmmmA}
\ee
are there, such that

\bigskip

\begin{enumerate}[label={${\text{Cond}}.\arabic*$}]
	
	\item\label{Cond1}
	Cycles $(m_1)_\infty$ and $(m_2)_\infty$ are disjoint (commute);
	
	\item	\label{Cond2}
	Cycle $(2)_1$ shares one element with $(m_1)_\infty$ and another with $(m_2)_\infty$;

	
	\item	\label{Cond3}
	Cycles $(m_1)_0$ and $(m_2)_0$ are disjoint (commute);
	
	\item	\label{Cond4}
	Cycle $(2)_u$ shares one element with $(m_1)_0$ and another with $(m_2)_0$.

\end{enumerate}

\bigskip

\noindent
One can fix the leftmost cycles as
\begin{align}
& (m_1)_\infty (m_2)_\infty (2)_1 	\nonumber
\\
	&= 
		(1, 2, \cdots,  m_1) 
		(m_1 + 1, m_1 +2, \cdots, m_1+m_2) 
		(1,m_1+1)		\nonumber
\\
	&= 
		(1, 2, \cdots,  m_1 ,
		m_1 + 1, m_1 +2, \cdots, m_1+m_2)		\label{m1m2cycl}
\end{align}
which is the most general form of satisfying \ref{Cond1}-\ref{Cond2}
modulo global $S_N$ transformations. (And now we cannot use $S_N$ transformations anymore.)
For example, with $m_1 = 4$ and $m_2 =3$, we fix
\begin{align}
(m_1)_\infty (m_2)_\infty (2)_1 
	&= 
		(1, 2, 3, 4)
		(5, 6, 7) 
		(1, 5) 	\nonumber
\\
	&= 
		(1, 2, 3, 4, 5, 6, 7) 	\label{Examcylc}
\end{align}
To satisfy Eq.(\ref{PermmmmA}), we the remaining cycles must be the inverse of (\ref{m1m2cycl}),
\begin{align}
( m_1 + m_2 , \cdots, 2, 1) = (2)_u (m_2)_0 (m_1)_0 .	\label{m1m2cyclin}
\end{align}
So our task reduces to counting in how many ways one can decompose the cycle in the l.h.s.~into a product of cycles with the structure in the r.h.s.~and satisfying \ref{Cond3}-\ref{Cond4}.

Our approach is to choose one element $k$ among the $m_1+m_2$ elements in the cycle in the l.h.s.~of (\ref{m1m2cyclin}) to be one of the two elements of $(2)_u \equiv (k, \ell)$. Once this is done, there are two ways of decomposing $(m_1+m_2, \cdots , k, \cdots, 1)$ according to the cyle structure in (\ref{m1m2cyclin}), namely

\begin{itemize}[$\cdot$]
	\item
	Choose the $m_2^{\rm{th}}$ element to the right of $k$ to be $\ell$; or
	
	\item
	Choose the $m_1^{\rm{th}}$ element to the right of $k$ to be $\ell$.

\end{itemize}
For example, choosing $k=6$ in the inverse of (\ref{Examcylc}) by marking it in green, the corresponding possible ways of fixing $\ell$ are marked in red:
\begin{align*}
(7,\textcolor{LimeGreen}{6},5,4,\textcolor{magenta}{3},2,1) 
	&= (\textcolor{LimeGreen}{6},\textcolor{magenta}{3}) (\textcolor{magenta}{3},5,4)(\textcolor{LimeGreen}{6},7,2,1)
\\
(7,\textcolor{LimeGreen}{6},5,4,3,\textcolor{magenta}{2},1) 
	&= (\textcolor{LimeGreen}{6},\textcolor{magenta}{2}) (\textcolor{LimeGreen}{6},1,7)(\textcolor{magenta}{2},5,4,3)
\end{align*}
Choosing next $k = 5$,
\begin{align*}
(7,6,\textcolor{LimeGreen}{5},4,3,\textcolor{magenta}{2},1) 
	&= (\textcolor{LimeGreen}{5},\textcolor{magenta}{2}) (\textcolor{magenta}{2},4,3)(\textcolor{LimeGreen}{5},1,7,6)
\\
(7,6,\textcolor{LimeGreen}{5},4,3,2,\textcolor{magenta}{1}) 
	&= (\textcolor{LimeGreen}{5},\textcolor{magenta}{1}) (\textcolor{LimeGreen}{5},7,6)(\textcolor{magenta}{1},4,3,2)	
\end{align*}

As we go on choosing the sites in $(m_1+m_2, \cdots ,1)$ one by one, once we arrive at $m_1$ sites away from the starting point, all possible decompositions have already  been found, and start repeating. In our example, the site at distance $m_1$ from 6 is 2, and the possible decompositions are
\begin{align*}
(7,\textcolor{magenta}{6},5,4,3,\textcolor{LimeGreen}{2},1) 
	&= (\textcolor{magenta}{6},\textcolor{LimeGreen}{2})(\textcolor{magenta}{6},1,7)(\textcolor{LimeGreen}{2},5,4,3)
\\
(7,6,\textcolor{magenta}{5},4,3,\textcolor{LimeGreen}{2},1) 
	&= (\textcolor{magenta}{5},\textcolor{LimeGreen}{2})(\textcolor{LimeGreen}{2},4,3)(\textcolor{magenta}{5},1,7,6)
\end{align*}
which we had already found before. In summary, we have found two different decompositions for each one out of $m_1$ elements (where $m_1 > m_2$). This proves (\ref{Hnumberm1m2}).





\bibliographystyle{utphys}

\bibliography{D1D5CompositeReferencesRepV16} 

\end{document}